\documentclass[twocolumn,preprintnumbers,amsmath,amssymb,showpacs,prb,aps]{revtex4}
\usepackage{txfonts}
\usepackage{pifont}
\usepackage{amssymb}
\usepackage{dcolumn}
\usepackage{amsmath}
\usepackage[dvips]{epsfig}
\usepackage {booktabs}

\makeatletter
\def\btt#1{\texttt{\@backslashchar#1}}%
\DeclareRobustCommand\bblash{\btt{\@backslashchar}}%
\makeatother

\begin{document}

\title{Band crossover and magnetic phase diagram of high-T$_C$ superconducting compound Ba$_{2}$CuO$_{4-\delta}$}

\author{Xiao-Cheng Bai$^{1,2,*}$}
\author{Ya-Min Quan$^{1}$}\emph{}
\thanks{These authors contributed equally. }
\author{H.-Q. Lin$^{3,4}$}
\author{Liang-Jian Zou$^{1,2}$}\emph{}
\thanks{zou@theory.issp.ac.cn}

\affiliation{ \it $^1$ Key Laboratory of Materials Physics,Institute of Solid State Physics, HFIPS,
	Chinese Academy of Sciences,Hefei 230031, China\\                
	\it $^2$ Science Island Branch of Graduate School,
	University of Science and Technology of China, Hefei 230026, China\\ 
\it $^3$ Beijing Computational Science Research Center, Beijing 100193, China \\
\it $^4$ Department of Physics, Beijing Normal University, Beijing 100875, China \\
}
\date{today}  

\begin{abstract}
We present the influences of electronic and magnetic correlations and doping evolution on the 
groundstate properties of recently discovered superconductor 
Ba$_{2}$CuO$_{4-\delta}$ by utilizing the Kotliar-Ruckenstein slave boson method.  
Starting with an effective two-orbital Hubbard model (Scalapino {\it et al.} Phys. Rev. {\bf B 99}, 224515 (2019)), 
we demonstrate that with increasing doping concentration, the paramagnetic (PM) system evolves 
from two-band character to single-band ones around the electron filling n=2.5, 
with the band nature of the  $d_{3z^{2}-r^{2}}$ and $d_{x^{2}-y^{2}}$ orbitals to
the $d_{x^{2}-y^{2}}$ orbital, slightly affected when the electronic correlation U varies from 2 to 4 eV.
Considering the magnetic correlations, the system displays one antiferromagnetically metallic (AFM) 
phase in $2<n<2.16$ and a PM phase in $n>2.16$ at U=2 eV, or two AFM phases in $2<n<2.57$ and 
$2.76<n<3$, and a PM phase in $2.57<n<2.76$
respectively, at U=4 eV. Our results show that near realistic superconducting state around n=2.6 
the intermediate correlated Ba$_{2}$CuO$_{3,2}$ should  be single band character,  and  the s-wave superconducting
 pairing strength becomes significant when U$>$2 eV, and crosses over to d-wave when U$>$2.2 eV.

\end{abstract} 
\maketitle

\section{\bf INTRODUCTION} 
\label{intro}

 % Multiorbital character:
 The discovery of newly cuprate superconductor Ba$_{2}$CuO$_{4-\delta}$ with  high Tc of 70 K \cite{JinCQ-PNAS-116-12156-2019}, 
about 2 to 3 times higher than isostructural conventional cuprate La$_{2}$CuO$_{4-\delta}$, have stirred  
great interesting on its essential electronic states and superconducting properties \cite{Maier-prb-99-224515-2019,
XiangT-prm-3-044802-2019,SZhou-prb-101-180509-2021,Hu-j-p-prb-99-224515-2019,wenzy-prm-3-044802-2020,
Aoik-prr-2-033356-2020,Oles-Orbital-Symmetry-2019,Sala-physicaC-2021} .
Though being isostructural to conventional cuprate La$_{2}$CuO$_{4-\delta}$, Ba$_{2}$CuO$_{4-\delta}$ 
displays compressed CuO$_6$ octahedra \cite{JinCQ-PNAS-116-12156-2019}, which are inverted to the CuO$_{6}$ 
octahadra or CuO$_{5}$ pyramids in the parent phases and their derivates of conventional cuprates
 La$_{2}$CuO$_{4}$, YBa$_{2}$Cu$_{3}$O$_{6}$ \cite{Grande-zaac-428-120-124-1977,Swinnea-jmr-2-424-426-1987}, etc.. 
This leads to 
unusual electronic states and superconducting nature \cite{Hu-j-p-prb-99-224515-2019}, in comparison with previous cuprates. 
 Recently Scalapino {\it et al.} proposed that the active orbiats near the Fermi 
energy in Ba$_{2}$CuO$_{4-\delta}$ are 3$d_{x^2-y^2}$ and 3$d_{3z^2-r^2}$ \cite{Maier-prb-99-224515-2019}. 
Such a two-orbital character is distinctly different from the single band nature of conventional cuprates \cite{Zhang-Rice-PRB-37-3759}.
Meanwhile, the effective single-band $t-J$ model was argued to be still applicable for the 
low-energy physics of the new cuprate. \cite{XiangT-prm-3-044802-2019,SZhou-prb-101-180509-2021}.
This debate immediately brings about the questions which scenario is applicable for superconducting 
Ba$_{2}$CuO$_{3.2}$, and which  composition, Ba$_{2}$CuO$_{3}$ or the Ba$_{2}$CuO$_{4}$, 
is the parent phase of superconducting Ba$_{2}$CuO$_{3.2}$.

Similar puzzle also may arise for earlier reported isostructural Sr$_{2}$CuO$_{4-\delta}$ 
with T$_c \approx$ 90 K \cite{JinCQ-PNAS-117-33099-2020}. 
Clarifying these questions is crucial for understand the superconducting nature of  Ba$_{2}$CuO$_{3.2}$.
% Coulomb correlation & OSMP
On the other hand, it is well known that conventional cuprates,  
La$_{2}$CuO$_{4-\delta}$, YBa$_{2}$Cu$_{3}$O$_{7-\delta}$, {\it etc.}, are strongly correlated 
systems with the Coulomb interaction considerably larger than the kinetic bandwidth. The strong 
correlation of Cu 3d electrons not only leads to well localized magnetic moment of Cu spins, but also
contributes non-Fermi liquid behavior and anomalous normal-state properties \cite{LeeP-RevModPhys-78-17}.  %PALee  
Presently it is not clear what role the electronic correlation plays in new superconductor in
 Ba$_{2}$CuO$_{4-\delta}$. Uncovering the roles of electronic correlations on the electronic states
is also crucial for understanding the ground state and pairing mechanism in Ba$_{2}$CuO$_{4-\delta}$, 
as well as in Sr$_{2}$CuO$_{4-\delta}$. 
In the same time, considering the possible multiorbital character in Ba$_{2}$CuO$_{4-\delta}$ \cite{Maier-prb-99-224515-2019}, 
one may naturally question whether Ba$_{2}$CuO$_{4-\delta}$ is in the orbital selective Mott 
phase (OSMP) when the electron correlation is strong enough \cite{Song-PhysRevB-103214510}.

Magnetic correlation has profound influences on the groundstate properties and 
magnetic structures of correlated systems. Previous works showed that 
spin correlations could considerably affect the magnetic groundstate and magnetic 
fluctuations of multiorbital Hubbard models \cite{Quan-EPJB-85-55-2012,Quan-CPC-191-90-2015,
Quan-APS-61-017016-2012,Quan-jmmm-456-329-332-2018,ZHANGyz-PRB-84-020401-2011,
ZHANGyz-PRB-85-035123-2012}.It remains unkown in recent found Ba$_{2}$CuO$_{4-\delta}$, 
what the groundstate magnetic structures of two parent compounds 
Ba$_{2}$CuO$_{3}$ and  Ba$_{2}$CuO$_{3.5}$ are? Clarifying these problems is important since as a 
new nonconventional superconductor, magnetic correlations and magnetic fluctuations are 
rather crucial for superconducting Cooper pairing.

In this work, we utilize the Kotliar-Runkenstein slave boson technique to study the two-orbital
Hubbard model of Ba$_{2}$CuO$_{4-\delta}$, focusing on
the evolutions of the electronic states with the increasing filling number and electronic correlation, 
as well as the evolutions of the Fermi surfaces in the normal state of in Ba$_{2}$CuO$_{4-\delta}$.
In Sec. II, we first describe the Kotliar-Runkenstein slave boson metheod. Then, in Sec. III, we discuss the numerical results.
We find that when the electron concentration increases, the system evolves from a two-band 
character to a single-band ones when the doping concentration increases, as well as two distinct antiferromagnetic 
phases at electron filling of $n=2$ and $3$, respectively.
To investigate the superconductivity in Ba$_{2}$CuO$_{4-\delta}$, 
in Sec. IV, we study the evolution of the pairing strength with the variations of the Coulomb correlation $U$ a
nd hole doping n within the random phase approximation.  The discussions and
conclusions are given in Sec. V.

\section {\bf Model Hamiltonian and Methods}
\label{model1}

We start from an effective two orbital Hubbard model at filling number $n=1+2\delta$ for Ba$_{2}$CuO$_{4-\delta}$ 
\cite{Maier-prb-99-224515-2019},
\begin{eqnarray}
\label{eq:Hamiltonian1}
H&=&H_{kin}+H_{loc}\\
H_{kin}&=&\sum_{i,j,\alpha,\beta,\sigma}    t^{\alpha\beta}_{ij}  c^{\dagger}_{i\alpha
   \sigma}c_{j\beta\sigma}+
   \sum_{i,\alpha,\sigma}\left(\varepsilon_{
   \alpha}-\mu\right)n_{i \alpha\sigma}  \\
H_{loc}&=&U\sum_{i,\alpha}n_{i\alpha\uparrow}n_{i\alpha\downarrow}
   +\sum_{i,\sigma,\sigma^{\prime},\alpha>\beta}
   \left(U^{\prime}-J_{H}\delta_{\sigma\sigma^{\prime}}\right)n_{i\alpha\sigma}
   n_{i\beta\sigma^{\prime}}               \nonumber\\
   &-&J_{X}\sum_{i,\alpha\neq\beta} c^{\dagger}_{i\alpha\uparrow}c_{i\alpha
   \downarrow}c^{\dagger}_{i\beta\downarrow}c_{i\beta\uparrow}
   +J_{P}\sum_{i,\alpha\neq\beta} c^{\dagger}_{i\alpha\uparrow}
   c^{\dagger}_{i\alpha\downarrow}c_{i\beta
   \downarrow}c_{i\beta\uparrow}
\end{eqnarray}
the $t^{\alpha\beta}_{ij}$ term describes electron hopping between the $i$-site with $\alpha$ orbital and the 
$j$-site with $\beta$ orbital.  
$c^{\dagger}_{i\alpha\sigma}$ ($c_{j\beta,\sigma}$) denotes the creation (annihilation)
operator of an electron with spin $\sigma=\uparrow(\downarrow)$ and orbital 
$\alpha(\beta)$. $n_{i\alpha\sigma}$ is the corresponding occupation number operator.
$\varepsilon_{\alpha} $ and $ \mu $ are the energy level and chemical potential. 
In the interaction Hamiltonian term, $H_{loc}$, the intra-band and inter-band Coulomb repulsions 
are denoted by $ U $ and $ U^{\prime} $. $ J_{H} $ , $ J_{X} $ and $ J_{P} $
are the Hund's rule coupling divided into Ising term, spin-flip term and pair-hopping term, respectively.
Here $ J_{X}=J_{P}=0 $ that stands for Ising Hund's coupling case is adopted in this paper.
Throughout this paper, we set $ U^{\prime}=U-2J_{H} $ and $ J_H=U/4 $. 
To investigate the ground state magnetic
properties of Ba$_{2}$CuO$_{4-\delta}$, we adopt the two-orbital tight-binding parameters 
given by T. Maier \textit{et al} for the two $e_{g}$ orbitals ($3d_{x^{2}-y^{2}}$,
$3d_{3z^{2}-r^{2}}$) near the Fermi level \cite{Maier-prb-99-224515-2019}.
The orbital-dependent dispersions $T^{\alpha\beta}$ in the Hamiltonian are as
follows \cite{Maier-prb-99-224515-2019},
 \begin{eqnarray}
\label{eq:dispersion}
T^{11/22}(\mathbf{k})&=&2t^{11/22}_{x/y}(\cos k_{x} +\cos k_{y})+
4t^{11/22}_{xy}\cos k_{x}\cos k_{y} \nonumber\\
& &+2t^{11/22}_{xx/yy}(\cos 2k_{x} +\cos 2k_{y}), \\
T^{12/21}(\mathbf{k})&=&2t^{12}_{x/y}(\cos k_{x}-\cos k_{y})+2t^{12}_{xx/yy}(\cos 2k_{x}-\cos 2k_{y}) \nonumber\\
\end{eqnarray}

The intra-orbital and inter-orbital hopping parameters
are shown in Table \ref{table1}. The on-site energy of orbital $d_{x^2-y^2}$ is $\varepsilon_{1}=-0.222$,
and the on-site energy of orbital $d_{3z^2-r^2}$ is $\varepsilon_{2}=\varepsilon_{1}+\Delta$,
where the crystal field splitting as a function of band filling $n$ is 
given by $\Delta=10.883-5n$ \cite{Maier-prb-99-224515-2019}.

\vspace{0.4cm}
\begin{table}[htb]
	\setlength{\arrayrulewidth}{0.8pt}
	\begin{center}
		\caption{The intra-orbital and inter-orbital hopping parameters
	 of the two-orbital tight-binding model in eV \cite{Maier-prb-99-224515-2019}.}
		\vspace{0.2cm}
		\begin{tabular}{|c|c|c|c|}
			\toprule[1.5pt]			
			\mbox{\rule[-1.2mm]{0.0mm}{5.0mm} No.}& $1^{st}$ ($t_{x/y}$) & $2^{nd}$ ($t_{xy}$)
			& $3^{rd}$ ($t_{xx/yy}$)   \\
			\hline
			intra-orbital $d_{x^{2}-y^{2}}$ & $t^{11}_{x/y}=-0.504$ & $t^{11}_{xy}=0.067$ & $t^{11}_{xx/yy}=-0.13$  
			\\
			\hline
			intra-orbital $d_{3z^{2}-r^{2}}$ & $t^{22}_{x/y}=-0.196$ & $t^{22}_{xy}=-0.026$ & $t^{22}_{xx/yy}=-0.029$ 
			\\
			\hline
			inter-orbital  & $t^{12}_{x/y}=0.302$ & $t^{12}_{xy}=0$ & $t^{12}_{xx/yy}=0.051$
			\\		
	\bottomrule[1.5pt]
			
		\end{tabular}
		\label{table1}
	\end{center}
\end{table}
\vspace{0.2cm}
  
\begin{figure}[htbp]
	\centering
	\includegraphics[angle=0, width=1.0 \columnwidth]{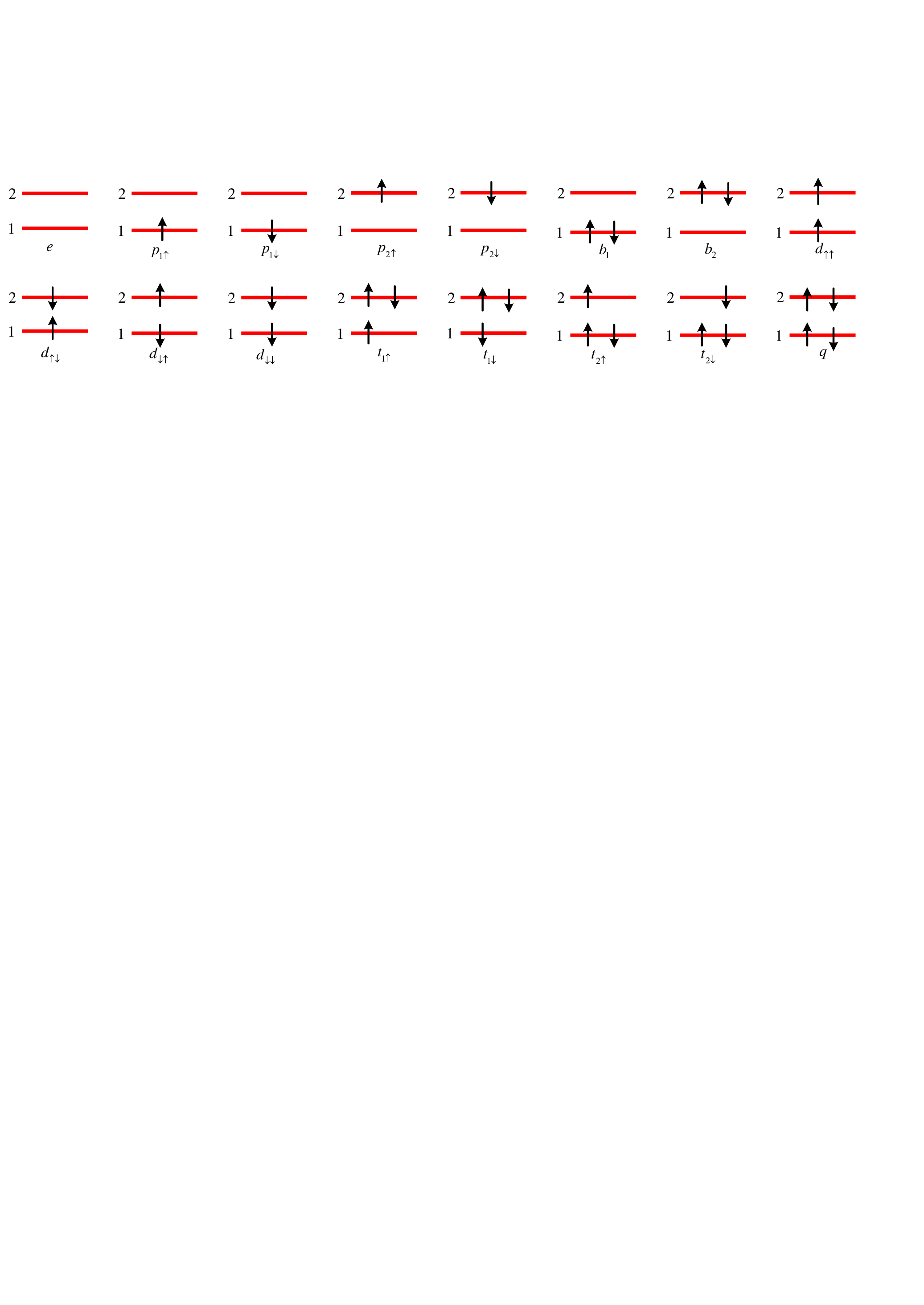}
	\caption{(Color online)The atomic spin and orbital configurations of two-orbital system
		 and the corresponding slave bosons. Here $e$, $p$, $b$, $d$, $t$, $q$ denote empty, singly, and
		 doubly occupied states in one orbital, doubly, triply and fully occupied states in different orbitals, respectively.}
	\label{fig:eigenstate}
\end{figure}

%CCCCCCCCCCCCCCCCCCCCCCCCCCCCCCCCCCCCCCCCCCCCCCCCCCCCC

 To investigate the magnetic groundstate properties of this model over wide electronic correlation, we use the
 Kotliar-Ruckenstein slave boson (KRSB) mean-field method 
 \cite{Kotliar-prl-57-1362-1986, Ruegg-epjb-48-55-2005, Hasegawa-prb-56-1196-1997}.
In the KRSB framwork, the local interaction term of the Hamiltonian
can be projected by slave-boson operators, thus the interaction Hamiltonian
term can be handled simply with saddle point approximation in the non-superconducting states.
To describe the multi-orbital Hubbard model in KRSB representation,
one should associate a boson creation operator $\phi^{\dagger}_{n}$ to every one of the
$2^{M}$ Fock states, where $M$ is the degree of each atom.
The saddle-point value of the slave boson is interpreted as the probability
of the corresponding Fock space configuration.
For a two-orbital KRSB model, the sixteen Fock states in each site are shown
in Fig.\ref{fig:eigenstate} and the corresponding slave boson operators
are also plotted. The sixteen slave bosons are classified into six categories 
according to the corresponding atomic configurations. We denote the slave boson 
operators $\phi^{\dagger}_{n}$ as $ \left\lbrace e^{(\dag)},p^{(\dag)}_{\alpha\sigma},
d^{(\dag)}_{\sigma_{\alpha}\sigma^{\prime}_{\beta}}, 
 b^{(\dag)}_{\alpha}, t^{(\dag)}_{\alpha\sigma}, q^{(\dag)}  \right\rbrace  $.
 Here $ e $ is the slave boson for the empty state, $p_{\alpha\sigma}  $
 for the singly occupied state in orbital $ \alpha $ with spin $\sigma$,
 $ d_{\sigma_{\alpha}\sigma^{\prime}_{\beta}} $ for the doubly occupied
 state with a $\sigma$-spin electron in the $\alpha$ orbital and a $ \sigma^{\prime} $-spin
 electron in the $\beta$ orbital. $ b_{\alpha} $ for the doubly occupied state with a 
 pair of up and down spin electrons in the $\alpha$ orbital. 
 $t_{\alpha\sigma}$ for the triply occupied state with a pair of electrons in the $\beta$ orbital
 and an extra $\sigma$-spin electron in the $\alpha$ orbital, $q$ for the fully occupied
 state, respectively \cite{Quan-EPJB-85-55-2012}.

 With these slave boson operators, the realistic electronic states are described 
 by these bosons and introduced auxiliary fermions $f^{\dagger}_{\alpha \sigma}$.
 Thus the realistic state $|\underline{n}\rangle$ represented in the enlarged Hilbert space, which is 
 the product of the slave bosons and the quasi-particle states, is 
 as follows \cite{Lechermann-PRB-76-155102-2007}: 
\begin{eqnarray}
\label{eq:fock1}
|\underline{n}\rangle\equiv \phi^{\dag}_{n}|Vac\rangle\otimes|n\rangle_{f},
\end{eqnarray} 
where the underline in $|\underline{n}\rangle$ denotes the representative
 state in the enlarged Hilbert space and $|n\rangle_{f}$ is the nth Fock state in
 the quasi-particle space. It is apparently not all the states in the enlarged Hilbert space 
 have physical significance. To enforce the solution in the physical subspace 
 in the enlarged Hilber space, all the slave-boson operators should satisfy
the following  normalization constraint and the fermion number constraint:

 \begin{eqnarray}
 \label{eq:constraint1}
 \sum_{n=1}^{2^{M}}\phi_{n}^{\dag}\phi_{n}=1 
 \end{eqnarray}
 and 
 \begin{eqnarray}
 \label{eq:constraint2}
 \hat{Q}_{\alpha\sigma}
(\phi)=f_{\alpha \sigma}^{\dag}f_{\alpha \sigma},  \qquad  \forall \alpha,
 \end{eqnarray} 
 where
 \begin{eqnarray} 
 \label{eq:slave_particle}
 \hat{Q}_{\alpha\sigma}
 (\phi)&\equiv& \sum_{n=1}^{2^{M}} \langle n|f_{\alpha \sigma}^{\dag}f_{\alpha \sigma}
 |n\rangle \phi_{n}^{\dag}\phi_{n}.  
 \end{eqnarray}

In the KRSB framework, the Eq.(\ref{eq:constraint1}) and Eq.(\ref{eq:constraint2})
 are rewritten as follows:  
 \begin{eqnarray}
 \label{eq:normalization}
 1&=&e_{}^{\dag}e_{}+\sum_{\alpha,\sigma}(p^{\dag}_{\alpha\sigma}p_{\alpha\sigma}
 +t^{\dag}_{\alpha\sigma}t_{\alpha\sigma})
 +\sum_{\alpha}b^{\dag}_{\alpha}b_{\alpha}\qquad \nonumber\\
 & &+\sum_{\sigma,\sigma^{\prime}}d^{\dag}_{\sigma_{\alpha}\sigma_{\beta}^{\prime}}
 d_{\sigma_{\alpha}\sigma_{\beta}^{\prime}}+q^{\dag}q,\\
 f^{\dag}_{\alpha\sigma}f_{\alpha\sigma}&=&p^{\dag}_{\alpha\sigma}p_{\alpha\sigma}
 +b^{\dag}_{\alpha}b_{\alpha}+\sum_{\sigma^{\prime}}d^{\dag}_{\sigma_{
 		\alpha}\sigma^{\prime}_{\beta}}d_{\sigma_{\alpha}\sigma^{\prime}_{\beta}}+\nonumber\\
 & &\sum_{\sigma^{\prime}}t^{\dag}_{\beta\sigma^{\prime}}t_{\beta\sigma^{\prime}}
 +t^{\dag}_{\alpha\sigma}t_{\alpha\sigma} +q^{\dag}_{}q_{}.   
 \end{eqnarray}
  
 The electron creation operator should display the same character in the origional Hilbert space and 
 the enlarged Hilbert space, that is $\langle \underline{m}|\underline{c}_{\alpha \sigma}^{\dag}|\underline{n}\rangle
 =\langle m|c_{\alpha \sigma}^{\dag}|n\rangle$, hence one has
  $\underline{c}_{\alpha \sigma}^{\dag}=
 \sum_{m n}\langle n|c_{\alpha \sigma}^{\dag}|m\rangle |\underline{n}\rangle \langle \underline{m}|$. 
 According to Eq.(\ref{eq:fock1}), the electron creation operator
  in the enlarged Hilbert space is definned as
  $\underline{c}_{\alpha \sigma}^{\dag}=
 \sum_{m n}  \langle n|f_{\alpha \sigma}^{\dag}|m\rangle \phi_{n}^{\dag}\phi_{m} 
 f_{\alpha \sigma}^{\dag}$ \cite{Lechermann-PRB-76-155102-2007}.
 To yield the noninteracting limit at saddle point approximation, a normalization
 term should be multiplied, thus the electron creation operator 
 in the slave boson representation takes the form \cite{Lechermann-PRB-76-155102-2007}: 
 \begin{eqnarray}
\label{eq:operator1}
\underline{c}_{\alpha \sigma}^{\dag}&=& f^{\dag}_{\alpha\sigma}Z^{\dag}_{\alpha\sigma},  
\end{eqnarray}
with
 \begin{eqnarray}
\label{eq:renor_factor1}
Z^{\dag}_{\alpha\sigma}&=&\sum_{m,n=1}^{2^{M}} \hat{Q}_{\alpha\sigma}^{-\frac{1}{2}} 
\langle n|f_{\alpha \sigma}^{\dag}|m\rangle \phi_{n}^{\dag}\phi_{m} 
(1-\hat{Q}_{\alpha\sigma})^{-\frac{1}{2}}
\end{eqnarray}

For a two-orbital system, the renormalization factor $Z_{\alpha\sigma}$ is given 
by \cite{Quan-EPJB-85-55-2012}
\begin{eqnarray}
\label{eq:renor_factor2}
Z_{\alpha\sigma} &=& \hat{Q}_{\alpha\sigma}^{-\frac{1}{2}}
\left(p^{\dag}_{\alpha\sigma}e+
b^{\dag}_{\alpha}p_{\alpha\bar{\sigma}}
+\sum_{\sigma'}d^{\dag}_{\sigma_{\alpha}\sigma'_{\beta}}p_{\beta\sigma'
}\right. \nonumber\\
& &\left. +t^{\dag}_{\alpha\sigma}b_{\beta}
+ \sum_{\sigma'}
t^{\dag}_{\beta\sigma'}d_{\bar{\sigma}_{\alpha}\sigma'_{\beta}}
+q^{\dag}t_{\alpha\bar{\sigma}}\right)\nonumber\\
& &(1-\hat{Q}_{\alpha\sigma})^{-\frac{1}{2}},
\end{eqnarray} 
 
%ccccccccccccccccccccccccccccccccccccccccccccccccccccccccc
The full Hamiltonian $H$ can be expressed in terms of the slave-boson and 
quasi-particle fermionic variables. The expression of the projected Hamiltonian in the enlargen Hilbert space
reads: 
\begin{eqnarray}
\label{eq:H_KRSB}
H&=&\sum_{\mathbf{k},\alpha,\beta,\sigma} T^{\alpha\beta}(\mathbf{k}) Z^{\dagger}_{\alpha
	\sigma}Z_{\beta,\sigma}  f^{\dagger}_{k\alpha
	\sigma}f^{}_{k\beta\sigma}+
\sum_{\mathbf{k},\alpha,\sigma}\left(\varepsilon_{
	\alpha}-\mu\right)f^{\dagger}_{k\alpha
	\sigma}f^{}_{k\alpha\sigma} \nonumber \\
&+&\sum_{n=1}^{2^{M}} \langle n|H_{loc}|n\rangle  \phi_{n}^{\dag}\phi_{n}.   
\end{eqnarray}

Within saddle-point approximation we can obtain the groundstate energy of the system. 
The mean-field Hamiltonian with constraints can be 
diagonalized, and the variational total energy of the system is given dy
\begin{eqnarray}
\label{eq:EGS_KRSB}
E_{tot}&=&\sum_{\mathbf{k},n} E_{n}(\mathbf{k})f(E_{n}(\mathbf{k}))+\sum_{n=1}^{2^{M}} \langle n|H_{loc}|n\rangle  \phi_{n}^{2} 
+\lambda \left( \sum_{n=1}^{2^{M}} \phi_{n}^{2}-1 \right) \nonumber\\
&-&\sum_{\alpha,\sigma} \eta_{\alpha \sigma} \left( Q_{\alpha\sigma}-n_{\alpha\sigma} \right),  
\end{eqnarray}
where $E_{n}$ is the eigenvalue of the Hamiltonian matrix and $f(E_{n}(\mathbf{k}))$ is the Fermi distribution function,
 $\lambda$ and $\eta_{\alpha \sigma}$ are the Lagrange multipliers used to implement the constraints, Eq.(\ref{eq:constraint1}) and 
 Eq.(\ref{eq:constraint2}), $n_{\alpha\sigma}$ is particle number 
of the orbital $\alpha$ with spin $\sigma$. The saddle-point equations are given by the partial 
derivatives of Eq.(\ref{eq:EGS_KRSB}) to all of the slave-boson amplitudes and
Lagrange multipliers. The self-consistent equations within saddle-point approximation are written as
follows:
 \begin{eqnarray}
 \label{eq:consistency} 
 \frac{\partial E_{gs}}{\partial \phi_{n}}&=&\sum_{\mathbf{k},n,\alpha,\sigma} 
 \frac{\partial E_{n}(\mathbf{k})}{\partial Z_{\alpha \sigma}}
 \frac{\partial Z_{\alpha \sigma}}{\partial \phi_{n}}+
 \sum_{n=1}^{2^{M}} 2\langle n|H_{loc}|n\rangle  \phi_{n}\nonumber\\ 
 & &+\lambda \sum_{n=1}^{2^{M}} 2\phi_{n}
 -\sum_{\alpha \sigma} \eta_{\alpha \sigma} \left( \frac{\partial Q_{\alpha\sigma}}{\partial \phi_{n}}
 -\frac{\partial n_{\alpha\sigma}}{\partial \phi_{n}} \right),\\ 
 \frac{\partial E_{gs}}{\partial \lambda}&=&\sum_{n=1}^{2^{M}} \phi_{n}^{2}-1,\\  
 \frac{\partial E_{gs}}{\partial \eta_{\alpha \sigma}}&=& Q_{\alpha\sigma}-n_{\alpha\sigma}   
 \end{eqnarray}

%ccccccccccccccccccccccccccccccccccccccccccccccccccccccccc

\begin{figure}[htbp]
	\centering
	\includegraphics[angle=0, width=0.45 \columnwidth]{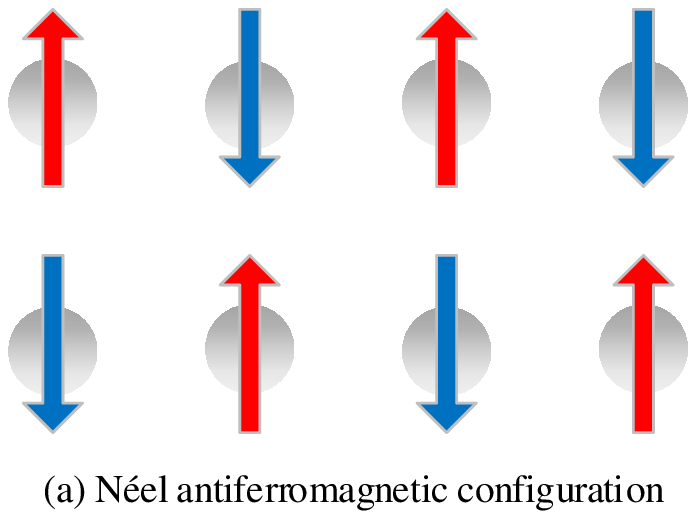}
	\includegraphics[angle=0, width=0.45 \columnwidth]{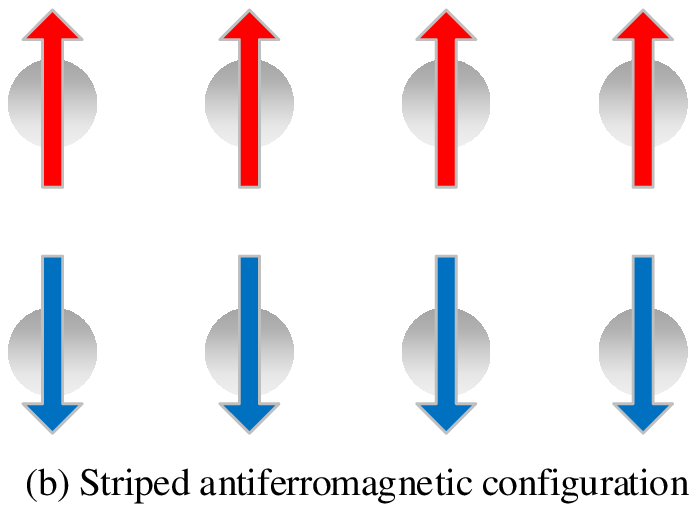}
	\caption{(Color online) Two possible spin configurations of the Cu spins of(a)N\'{e}el AF, (b)Striped AF.
		The gray balls represent Cu atoms, the red and blue arrows denote the spin directions.}
	\label{fig:spin_configuration}
\end{figure}

We get the ground state by solving the self-consistent equations through optimizing 
the objective function defined as \cite{Quan-CPC-191-90-2015}
% from the self-consistent equations \cite{Quan-CPC-191-90-2015,Quan-APS-61-017016-2012,Quan-jmmm-456-329-332-2018}
\begin{eqnarray}
\label{eq:consistency}
F\left( \phi_{n},\lambda,\eta_{\alpha \sigma} \right)=
\sum_{n=1}^{2^{M}}  \left( \frac{\partial E_{gs}}{\partial \phi_{n}}  \right)^{2}+
\left( \frac{\partial E_{gs}}{\partial \lambda}  \right)^{2}+
\sum_{\alpha \sigma} \left(\frac{\partial E_{gs}}{\partial \eta_{\alpha \sigma}} \right)^{2}. 
\end{eqnarray}

Throughout this paper we compare the total energies of four possible candidates, paramagnetic metallic, 
ferromagnetic, N\'{e}el antiferromagnetic and striped antiferromagnetic phases, 
so as to find the most stable phase as the ground state. The latter two phases are shown in Fig.2.

To investigate the superconductivity in Ba$_{2}$CuO$_{4-\delta}$, based on the original Hamiltonian in Eqns.(1-5), 
we study the evolution of the pairing strength with the variations of the Coulomb correlation $U$ and hole doping n
through the random phase approximation. The pairing strength  $\lambda^{\alpha}$ ( $\alpha$=s, d and g) 
could be obtained by solving the linearized gap 
equation \cite{Maier-prb-99-224515-2019}
\begin{eqnarray}
\label{eq:consistency}
-\sum_{j} \oint \frac{dk'_{||}}{2\pi v_{F_{j}}\left(k'_{||}\right)}\Gamma_{i,j}\left(k,k'\right) 
g_{j}^{\alpha}\left(k'\right)=\lambda_{\alpha} g_{i}^{\alpha}\left(k\right),
\end{eqnarray}
Here, $\alpha$ is the pairing channel, $i$ and $j$ are the band indices of Fermi surface vectors $k$, $k'$, respectively,   
$v_{F_{j}}\left( k'_{||} \right)$ is the Fermi velocity, and $\Gamma_{i,j}\left(k,k'\right)$
is the superconducting pairing interaction. With these equation we could well reproduce the results in Ref.[2].

%%ccccccccccccccccccccccccccccccccccccccccccccccccccccccccccccccccccccccccc
\section{NUMERICAL RESULTS}
\label{results}

To demonstrate the effects of doping and electronic correlation on the electronic states,
we present the numerical results of evolution of electronic states with the incrasing U in
the electron filling range of $2<n<3$. 
We find that the ferromagnetic state is always high energy, hence is neglected in what follows.

\subsection{\bf Paramagnetic Phases} 
 
      We first present the evolutions of the band structures and the Fermi surfaces of 
paramagnetic Ba$_2$CuO$_{4-\delta}$ with increasing doping for U=2 and 4 eV, as shown in Fig.3 and Fig.4 respectively. 
 \begin{figure}[htbp]
	\centering
	\includegraphics[angle=0, width=0.49 \columnwidth]{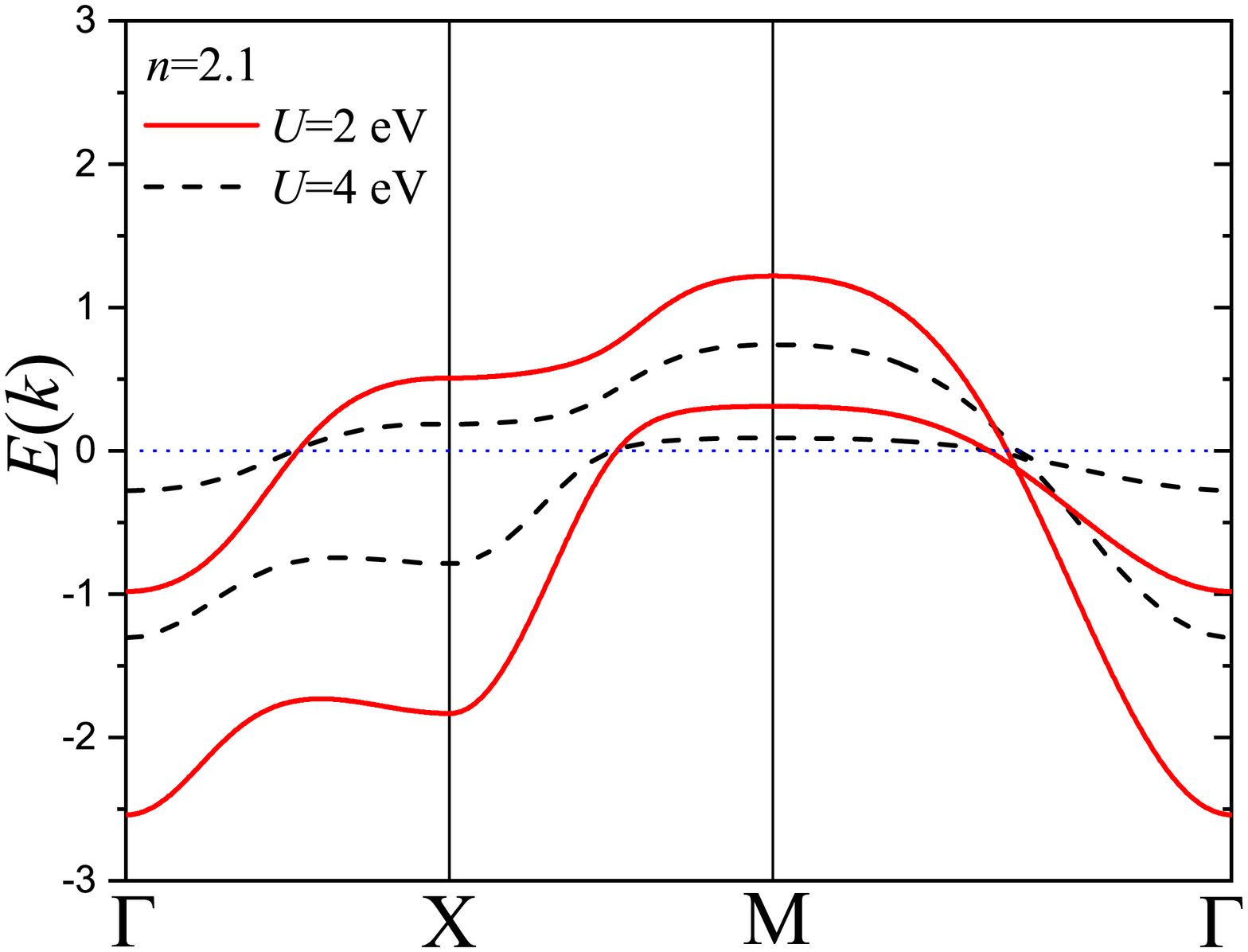}
	\includegraphics[angle=0, width=0.49 \columnwidth]{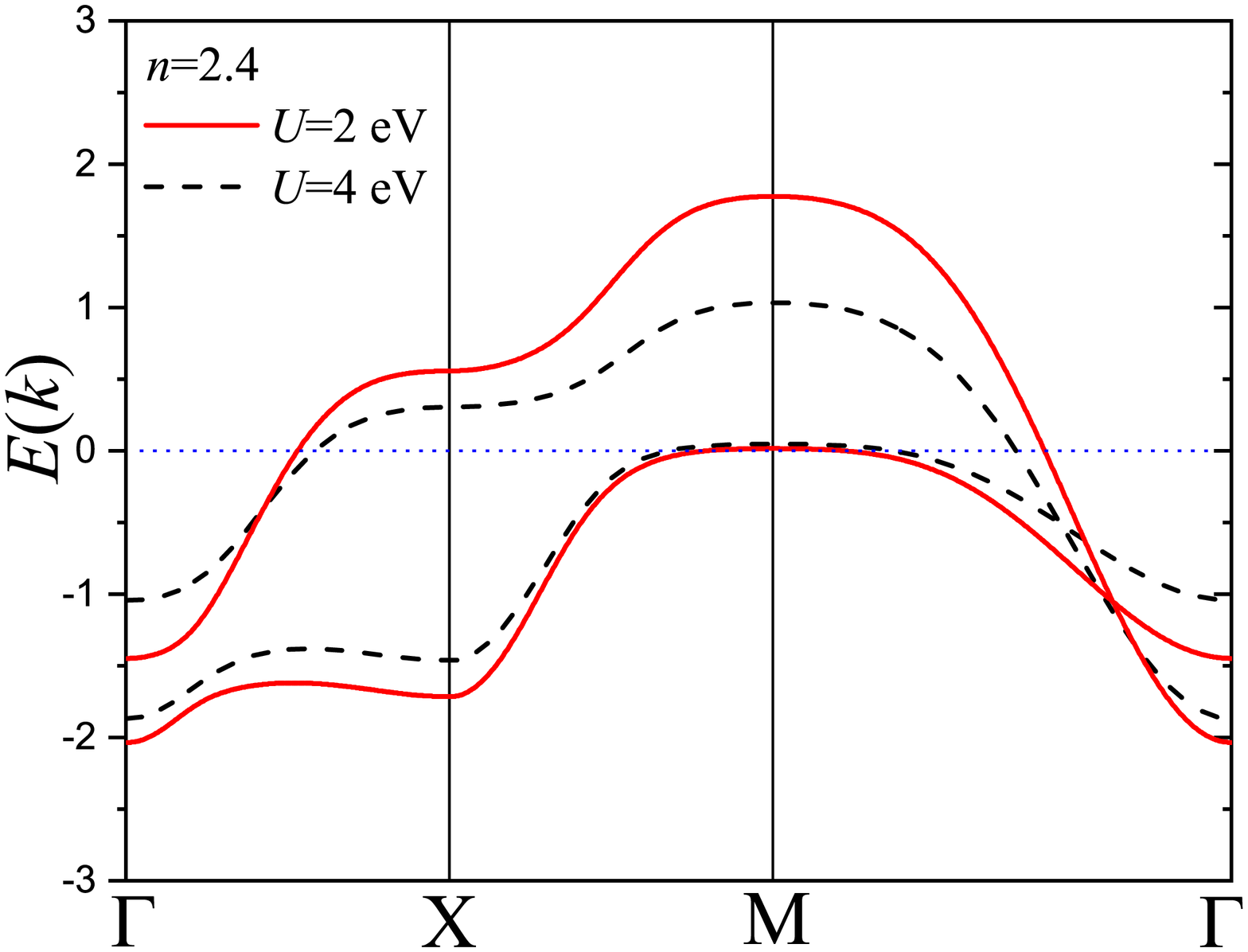}
	\includegraphics[angle=0, width=0.49 \columnwidth]{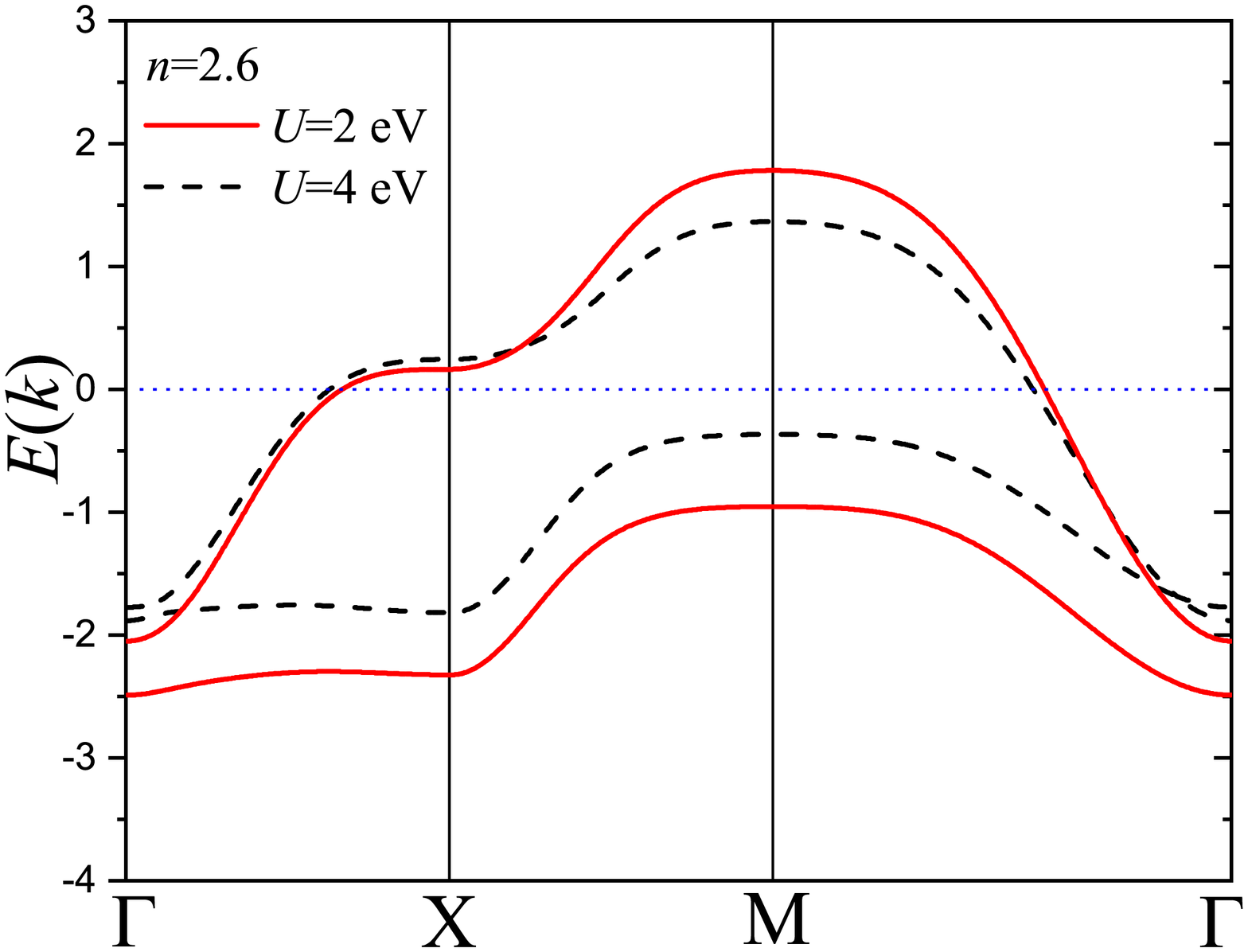}
    \includegraphics[angle=0, width=0.49 \columnwidth]{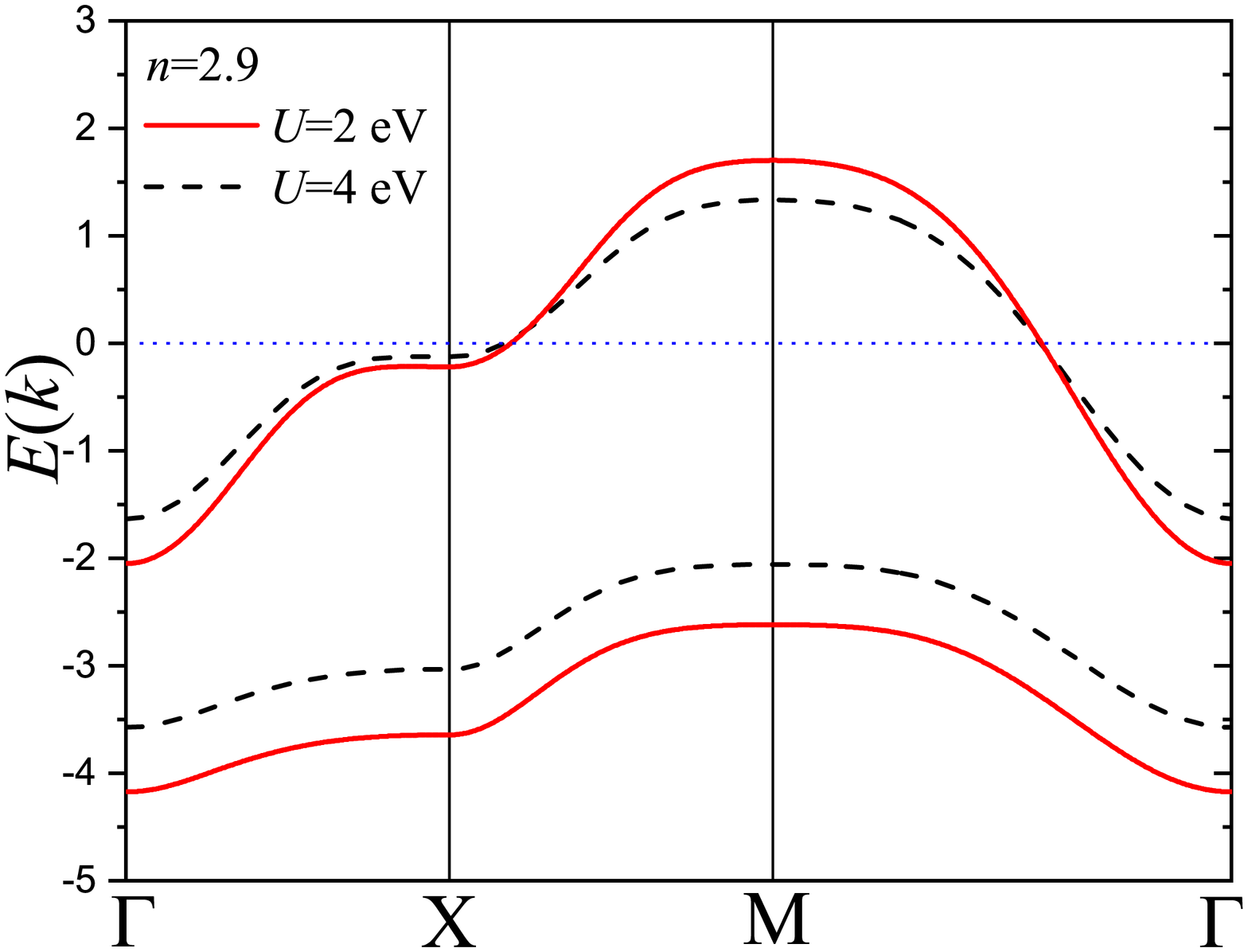} 
	\caption{(Color online) Dependence of the band structures on different doping in paramagnetic 
	Ba$_{2}$CuO$_{4-\delta}$ at $U=2$ (dotline) and $U=4$ (dsahline).  } 	   
	\label{Fig.103}
\end{figure} 
We find that both for the intermediate correlation of U=2 eV and the strong correlation of U=4 eV,
the systems are two bands near E$_F$ for the filling number n=2.1 and 2.4, and cross over to 
single band near E$_F$ for the filling number n=2.6 and n=2.9, as in Fig.3. 
According to the analysis to the orbital weight of two-band situations and the single band situations, 
as seen the {\it supplementary material},
one finds that the two bands consist of the admixture of the 3d$_{3z^2-r^2}$ and 3d$_{x^2-y^2}$
orbitals due to considerable interorbital hopping t$^{12}$, and the single band mainly contributes 
from the 3d$_{3z^2-r^2}$ orbital. For optimized doped superconducting phase Ba$_{2}$CuO$_{3.2}$, 
$n \approx 2.6$, only one correlation band crossing Fermi level suggests
that Ba$_2$CuO$_{3.2}$ should be essentially a single band superconductor.

The doping evolution of the Fermi surfaces shown in Fig.4 further demonstrates the crossover 
character of two band scenario to single band one. Fig.4 shows that in the PM phase the electronic correlation does not change the
Fermi surfaces too much, the doping drives the system crossover from two-type Fermi surfaces with electron
and hole carriers to single hole Fermi surface,
leading to the {\it Lifshitz} transition in Ba$_{2}$CuO$_{4-\delta}$.
 \begin{figure}[htbp]
	\centering
	\includegraphics[angle=0, width=0.49 \columnwidth]{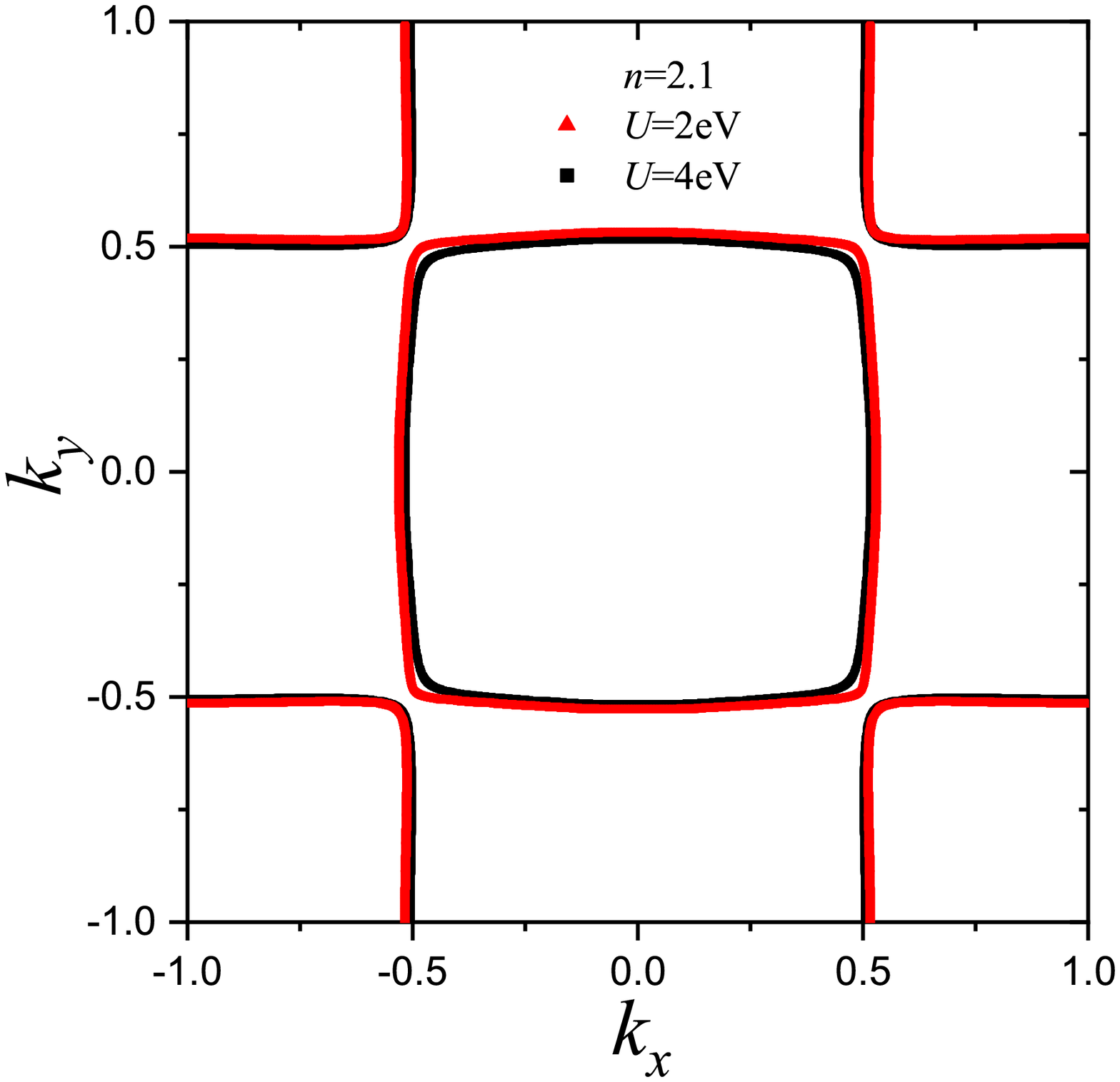}
	\includegraphics[angle=0, width=0.49 \columnwidth]{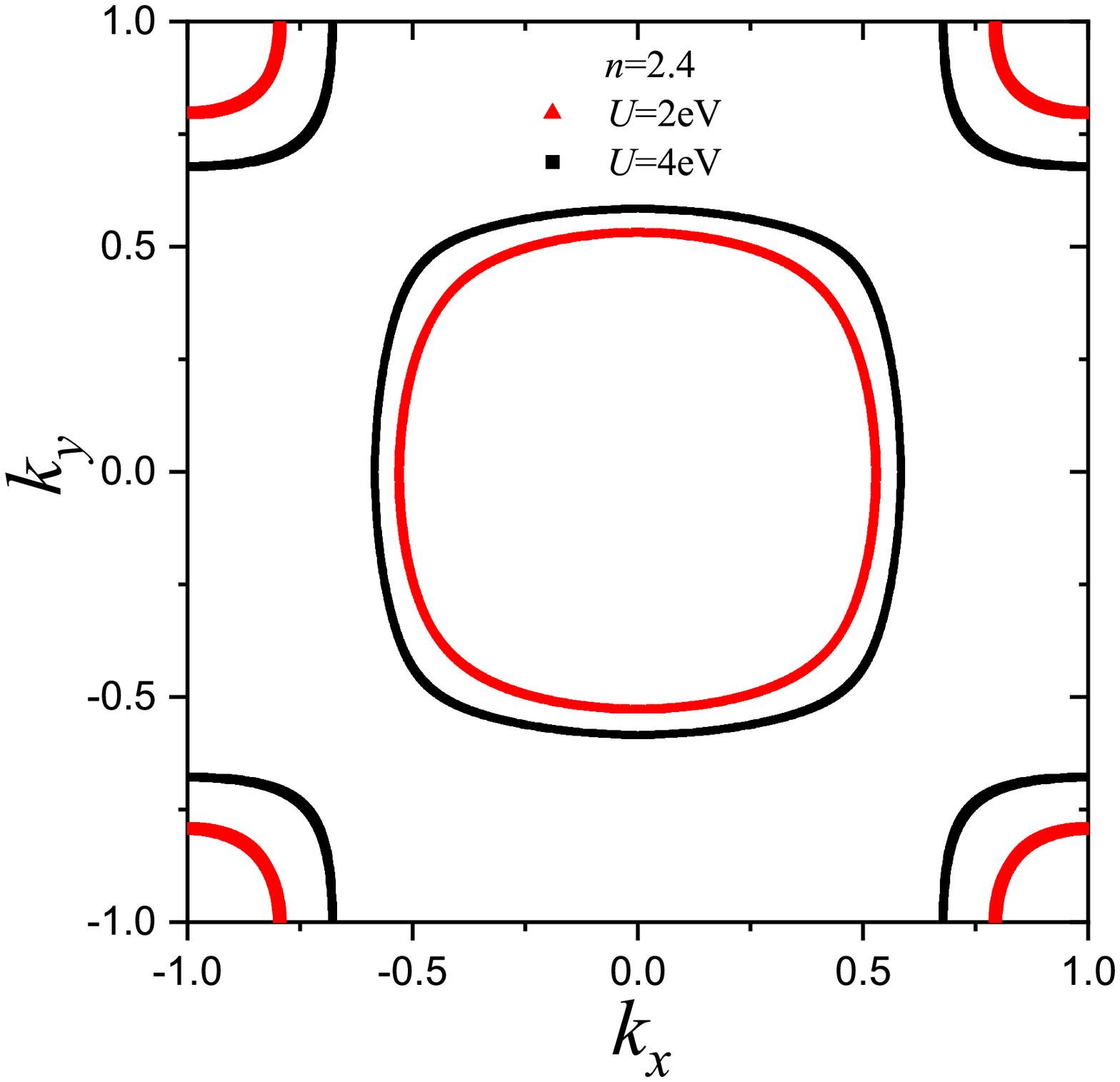}
	\includegraphics[angle=0, width=0.49 \columnwidth]{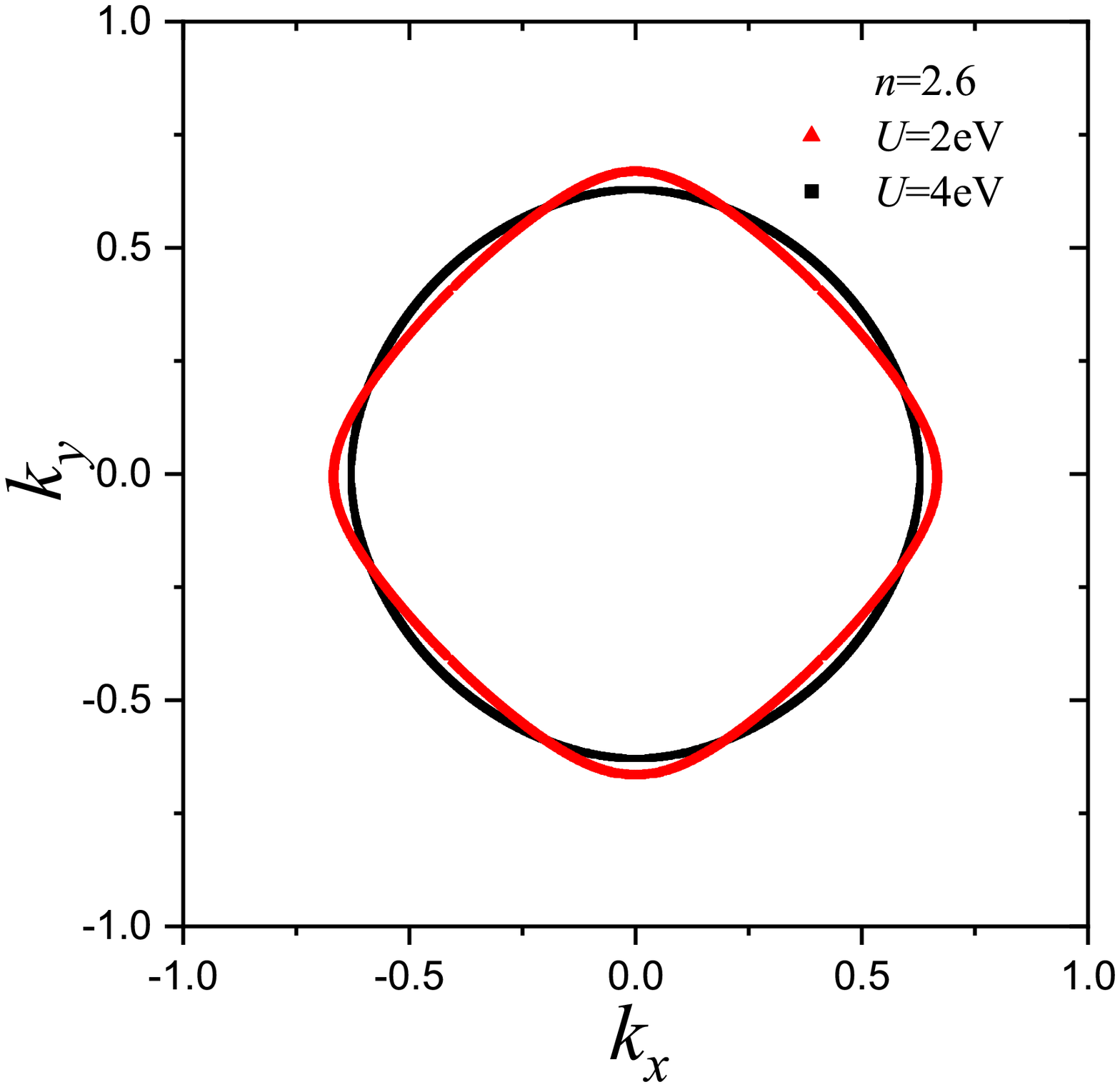}
	\includegraphics[angle=0, width=0.49 \columnwidth]{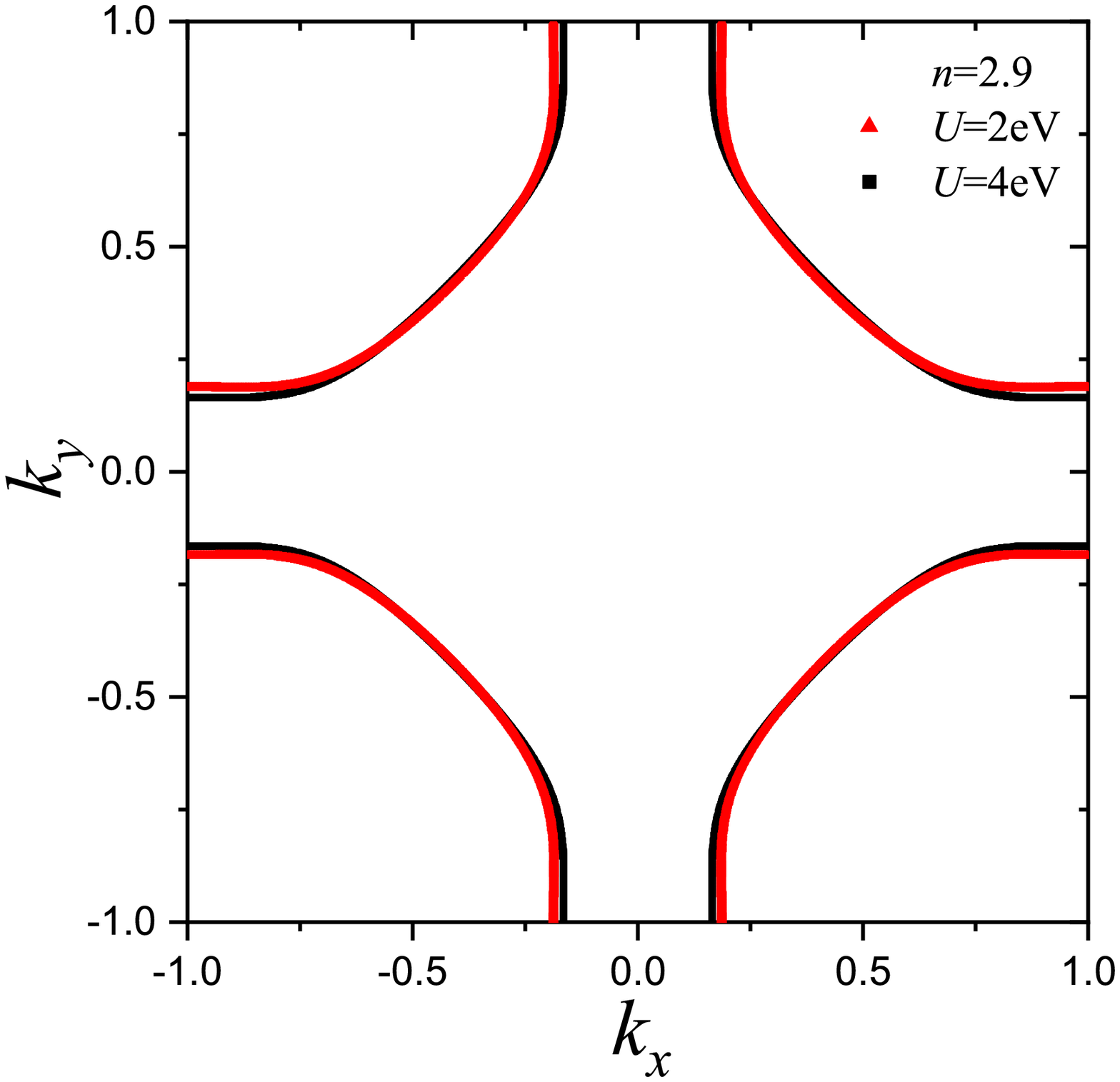} 
	\caption{(Color online) The Fermi surfaces of Ba$_{2}$CuO$_{4-\delta}$
		for different doping at $U=2$  and $U=4$. } 	   
	\label{Fig.104}
\end{figure} 
The orbital weights of these Fermi surfaces are in agreement with those of the bands.

\subsection{\bf  Magnetically Ordered Phases}

In the presence of electronic correlation, the paramagnetic phase is usually unstable 
regarding  to magnetically ordered phases. In this subsection we present the magnetic
phase diagrams of Ba$_2$CuO$_{4-\delta}$ upon the increases of the electron correlation U
and the doping concentration. We first present two integer-filling compounds with n=2 and 3, 
which are two possible parent phases of superconducting states, and then the general doping case.

\subsubsection{\bf Half-filling Case}
 
 Firstly, we present the magnetic phase diagram of Ba$_2$CuO$_{4-\delta}$ on the correlation dependence 
at $n$=2 in Fig.5. This filling corresponds to one possible parent phase Ba$_2$CuO$_{3.5}$.
 \begin{figure}[htbp] 
 	\centering
 	\includegraphics[angle=0, width=1.00 \columnwidth]{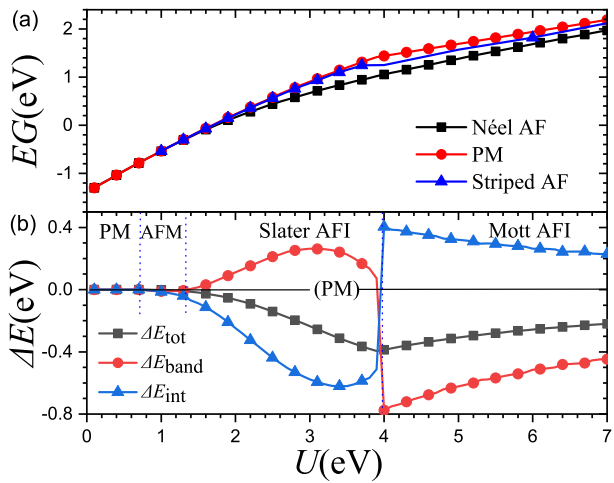}
 	\caption{(Color online) $U$ dependence of the total energy $E_{tot}$ of three different magnetic 
	         phases (a)  and of the total, band and interaction energy differences  $\Delta E_{tot}$, 
	         $\Delta E_{band}$ and $\Delta E_{int}$ between the N\'{e}el AF and PM phases. 
	         Here PM, AFM and AFI denote the paramagnetic metallic phase, N\'{e}el antiferromagnetic 
		 metallic and insulating phases, respectively. Theoretical parameters: n=2, $J_{H}=U/4 $}
 	\label{Fig.13}
 \end{figure} 
 The $U$-dependence of the total energies of the PM, the N\'{e}el AF and the striped 
 AF states are shown in Fig.\ref{Fig.13}(a). It shows that when $U<0.65$ eV, these states are almost degenerate;
 when $U>0.65$ eV, the total energy of the N\'{e}el AF state is considerably lower than the other two, 
 suggesting that the N\'{e}el AF state is the most stable.

Analyzing the contributions of the kinetic or band energy and the potential or interaction energy to the total energy, 
we disclose the phase competition more detail, as shown in Fig.\ref{Fig.13}(b). 
 The total energy difference in Fig.\ref{Fig.13}(b) is defined as $\Delta E_{tot}=E_{Neel}-E_{PM}$, and
 $\Delta E_{band}$ and $\Delta E_{int}$ denote the band and interaction energy differences between the 
 N\'{e}el AF and PM phases, respectively. 
It can be seen that with the increasing electronic correlation, the system sequently transits from the 
PM to the N\'{e}el AF metal, to the Slater AF insulator (AFI), and to the Mott 
AFI at $U_c$=0.65, 1.3, 4 eV, respectively.  
Here the Slater AFI phase is driven by interaction energy, so $ \Delta E_{int}<0$ and $\Delta E_{band}>0$; on
the contrast, the Mott AFI phase is driven by kinematic energy, so
 $\Delta E_{int}>0$ and $\Delta E_{band}<0$ \cite{Watanabe-prb-89-165115-2014}. 
In the magnetically insulator the total energy difference between the PM and N\'{e}el AF states gives rise to the magnetic 
energy difference, or the spin coupling strength. From Fig.5(b) one estimates that  magnetic 
energy differences are about 100, 150, 230 and 400 meV for U=2, 2.5, 3 and 4 eV, respectively. 
These results demonstrate that when the system is in the intermediate correlation regime with $U=2$ eV, the ground state is
a Slater AFI; when it is in the strong correlation regime with $U>4$ eV, the ground state is a Mott AFI.

\subsubsection{\bf Three-quarter filling case}
 
Then we present the magnetic phase diagram of Ba$_2$CuO$_{4-\delta}$ on the correlation dependence at n=3, which is 
another possible parent phase of the superconducting state. The two-orbital model with the electron filling $n=3$ 
 describes the compound Ba$_{2}$CuO$_{3}$. The U-dependences of the total energies in the PM, N\'{e}el AF, 
 and striped AF configurations and of the energy differences are similar to Fig.5, except for the 
 phase boundaries, as seen in Fig.6.
 \begin{figure}[htbp]
 	\centering
 	\includegraphics[angle=0, width=1.00 \columnwidth]{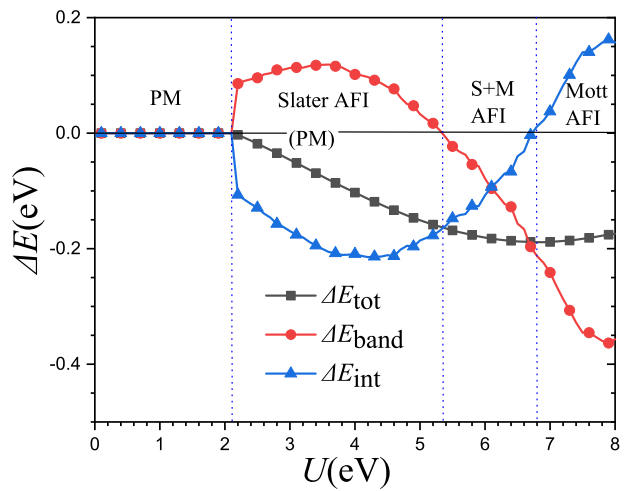}
 	\caption{(Color online) The $U$ dependence of total energies of Ba$_2$CuO$_{4-\delta}$ in the PM, and N\'{e}el AF 
	configurations (a) and  of the energy differences $\Delta E_{tot}$,$\Delta E_{band}$,	$\Delta E_{int}$ at n=3. 
	        Other parameters are the same to Fig.5.}
 	\label{Fig.21}
 \end{figure}
 One finds that when the electron correlation increases from 0 to 8 eV, the sequently quantum phase transitions 
 from the PM to Slater AFI, and Slater AFI to Mott AFI happen at 2.1 and 6.8 eV, respectively.
 Comparing with the results at $n=2$, we find that more large critical values $U_c$ are needed to drive the PM-Slater AFI 
 and the Slater-Mott AFI transitions.
This demonstrate that when the system is in the intermediate correlation regime with $U<2.1$ eV, the ground state is
a PM; when it is in the strong correlation regime with $U>$4 eV, the ground state is a Slater or Mott AFI.
Meanwhile one notices that in Fig.\ref{Fig.21}, the transition of the system from the PM to Slater AFI phases at $U=2.1$ eV 
is the first order, this arises from the effect of the crystalline field splliting.

%CCCCCCCCCCCCCCCCCCCCCCCCCCCCCCCCCCCCCCCCCCCCCCCCCCCCCCCCCCCCCCC

\subsubsection{\bf General Doping Cases}
 
To explore the doping evolution and the electron correlation effect in Ba$_2$CuO$_{4-\delta}$, 
we obtain the general doping dependences of the magnetic phase diagrams, sublattice magnetic moments, 
and the Fermi surfaces at different correlation strengths of U=2 eV and 4 eV, respectively. 
\\

\paragraph{\bf  Magnetic Phase Diagrams}

\begin{figure}[htbp]
	\centering
	\includegraphics[angle=0, width=1.00 \columnwidth]{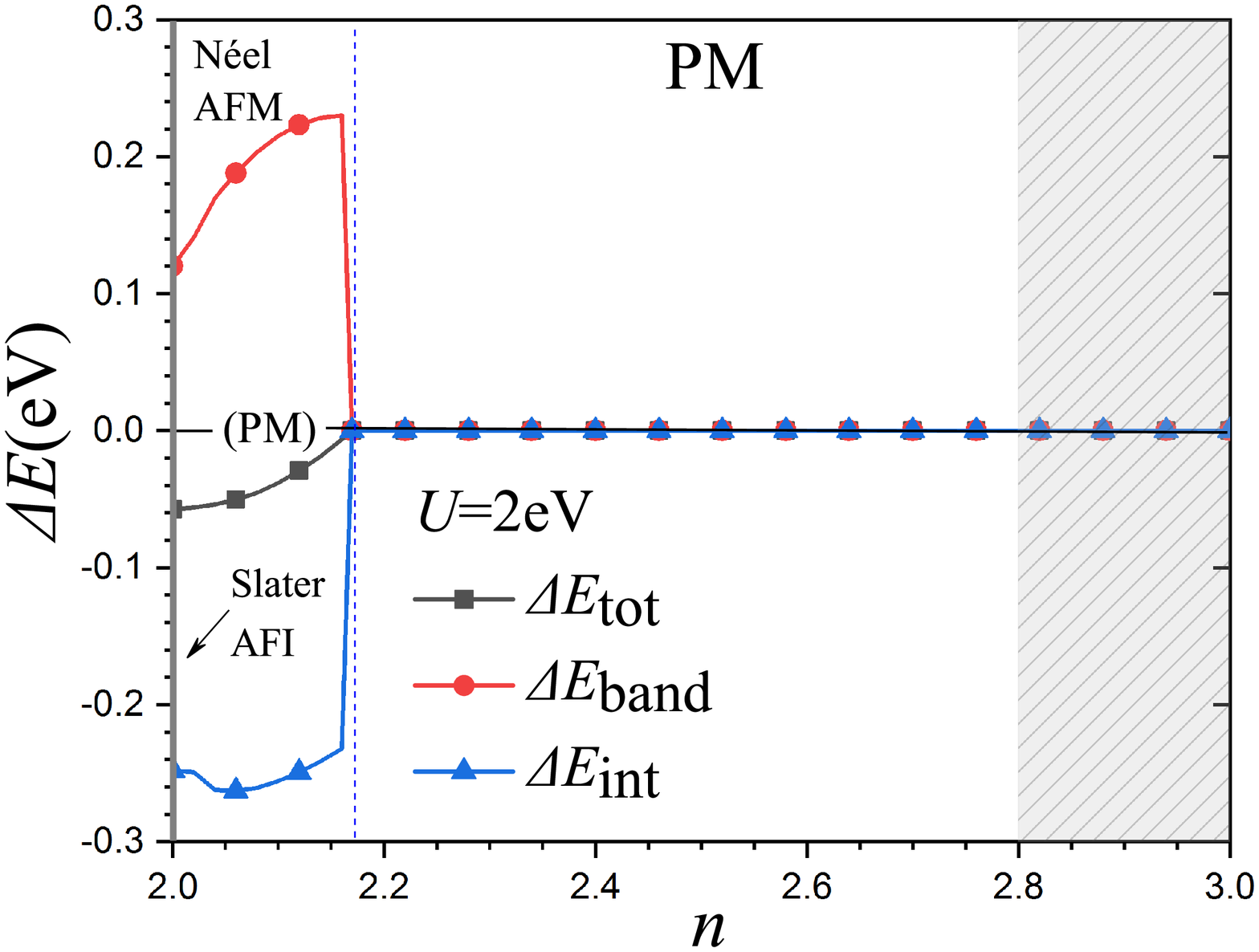}
	\includegraphics[angle=0, width=1.00 \columnwidth]{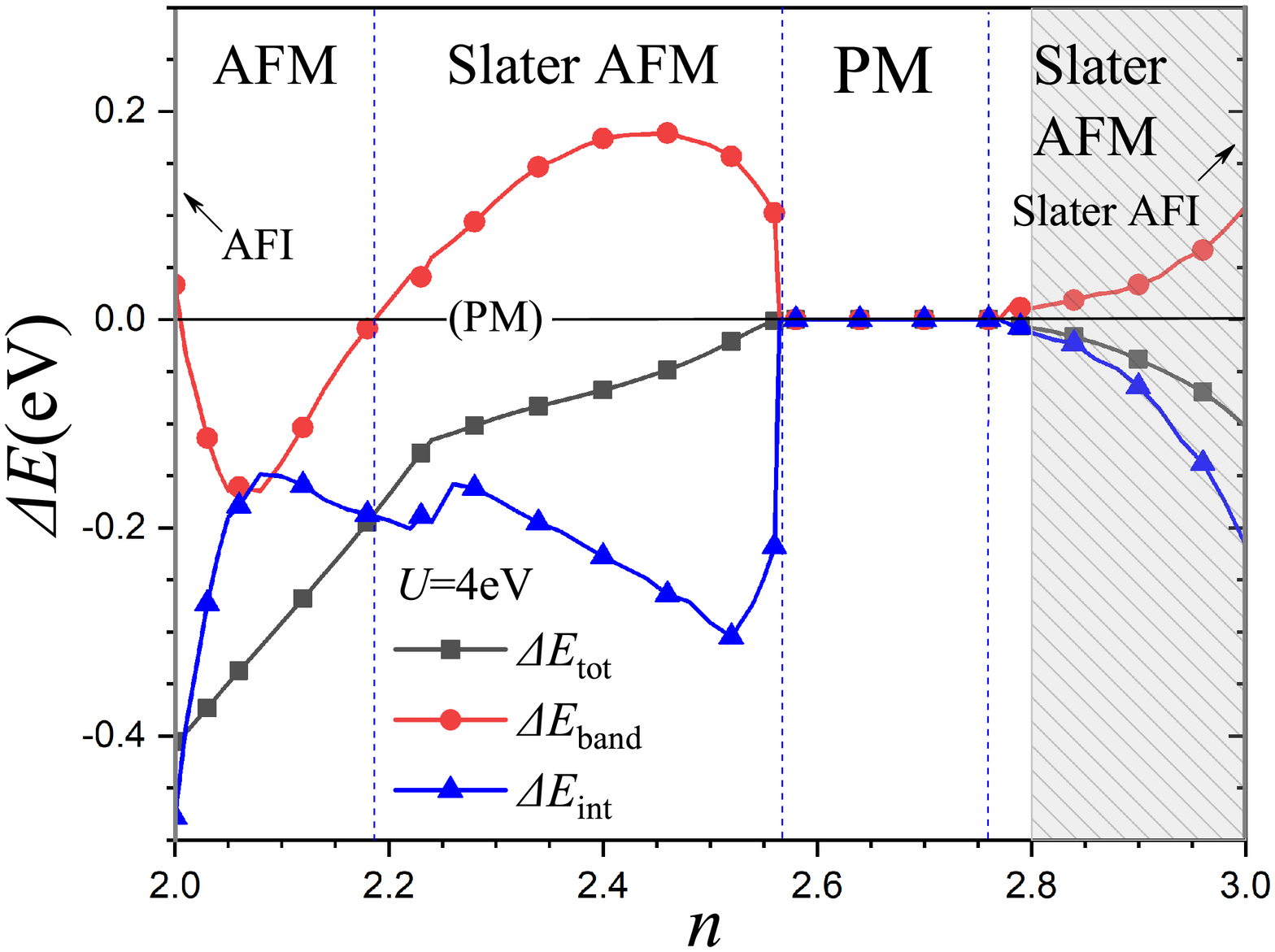}	
	\caption{(Color online)  The  magnetic phase diagrams of Ba$_2$CuO$_{4-\delta}$ and the energy differences 
	        between the PM state and the N\'{e}el AF state at U=2 eV (a) and U=4 eV (b).
		 } 	   
	\label{Fig.41}
\end{figure}

After comparing the total energies of various magnetic configurations and determining the ground states 
for various doping, we plot the magnetic phase diagrams in the particle number range $2<n<3$ in 
Fig.\ref{Fig.41} at U=2 and 4 eV,
respectively. Among various magnetic configurations, the PM, N\'{e}el  and striped   
AF and ferromagnetic phase are taken into account. The ferromagnetic and striped  
AF phases are always unstable when $U<4$, hence are neglected. So only the energy difference of  
the N\'{e}el AF state respect to the PM phase is plotted in Fig.\ref{Fig.41}.
  
One finds that in the intermediate correlation situation of U=2 eV shown in Fig.\ref{Fig.41}(a), the system is  
the N\'{e}el AF metallic (AFM) phase in the filling factor range of $2<n<2.18$; the system becomes the PM
in the wide range of $2.18<n<3$. The dashed region denotes dominant single-band/orbital range.
Such a simple magnetic phase diagram suggests that intermediate 
correlated Ba$_2$CuO$_{4-\delta}$ could not have two superconducting phases.

In the strongly correlated regime of U=4 eV shown in Fig.\ref{Fig.41}(b), the magnetic phase diagram 
of the system is rich. With the increase of filling number n, 
the system first evolves from Mott AFI at n=2 to the N\'{e}el AFM phase in the filling factor range 
of $2<n<2.18$, where the energy gain $\Delta E_{int}$ and $\Delta E_{band}$ are both negative.
This indicates that the AFM phase is interaction-energy and band-energy driven. 
With the further increasing $n$, only the energy gain $\Delta E_{int}$ are negative, the ground state of the system
enters to Slater AFM phase due to the increasing Columb screening effect in the filling number range of 
$2.18<n<2.56$.  
Furthermore the system enters the PM state in $2.56<n<2.75$. 
One notices that the system transits from the AFM  metallic to PM phase sharply at $n=2.56$.
This arises from the effect of the crystal field splitting,
which leads to a first order transition.
Further increasing n drives the system to another Slater AFM phase adjacent to the 
Slater AFI phase at n=3. In this range of $2.75<n<3.0$, the  $d_{3z^{2}-r^{2}}$ orbital is closely
fully occupied. The inter-orbital charge fluctuations and spin correlations are weak;
thus the system becomes an effective single band model in the vicinity of three-quarter filling, 
as shown the shaded region in Fig.\ref{Fig.41}(b), the energy gain of the AFM state is interaction-energy driven at $U=4$.
\\

\paragraph{\bf Sublattice Magnetic Moments}

\begin{figure}[htbp]
	\centering
	\includegraphics[angle=0, width=1.00 \columnwidth]{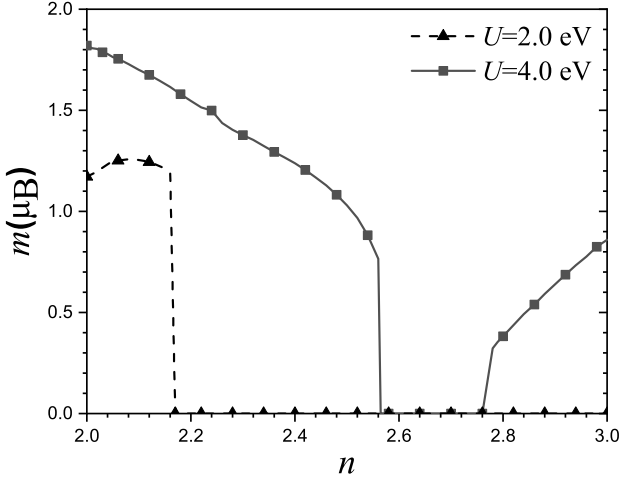} 
	\caption{(Color online) The sublattice magnetic moments of Ba$_2$CuO$_{4-\delta}$ at $U=2$ eV 
	              (black line)  and $U=4$ eV (red line)  as the functions of band filling . }
	\label{Fig.42}
\end{figure}

The  evolution of electronic states of Ba$_2$CuO$_{4-\delta}$ is also reflected in the doping
dependence of the sublattice magnetic moment of Cu spins at inequivalent sites. Fig.\ref{Fig.42}
displays the evolution of the sublattice magnetic moment with increasing filling number n for 
different correlation strength of U=2 eV and 4 eV. In the intermediate correlation regime at U=2 eV,
the magnetic moment, as seen the dashed line in Fig.\ref{Fig.42}, is about 1.2 $\mu_{B}$ in the AFM phase
when $2<n<2.18$. In this region
the active bands around Fermi level have two since $m>1$ $\mu_{B}$.
%CCCCCCCCCCCCCCCCCCCCCCCCCCCCCCCCCCCCCCCCCCCCCCCCCCCC 
%%ccccccccccccccccccccccccccccccccccccccccccccccccccccccccccccccccccccccccc
In strongly correlated Ba$_2$CuO$_{4-\delta}$ at $U=4$ eV, as seen the solid line in Fig.\ref{Fig.42}, 
the sublattice magnetic moment of Cu spins monotonically decreases from about 1.8 $\mu_B$ 
at n=2 to about 0.7 $\mu_B$ and diminishes to zero at n=2.56 in the first AFM region; Obviously,
the compound in this region is a typical two-band system. In the second AFM region, 
the sublattice magnetic moment increases from zero to about 0.85 $\mu_B$ when the filling number n
increases from  2.68 to 3. This supports that the compound in this region is a single band system. 
Detail analysis shows that the active band
is fully contributed by the electrons from the $d_{x^{2}-y^{2}}$ orbital. 
\\

\paragraph{\bf Fermi Surfaces}

\begin{figure}[htbp]
 	\centering
 	\includegraphics[angle=0, width=0.49 \columnwidth]{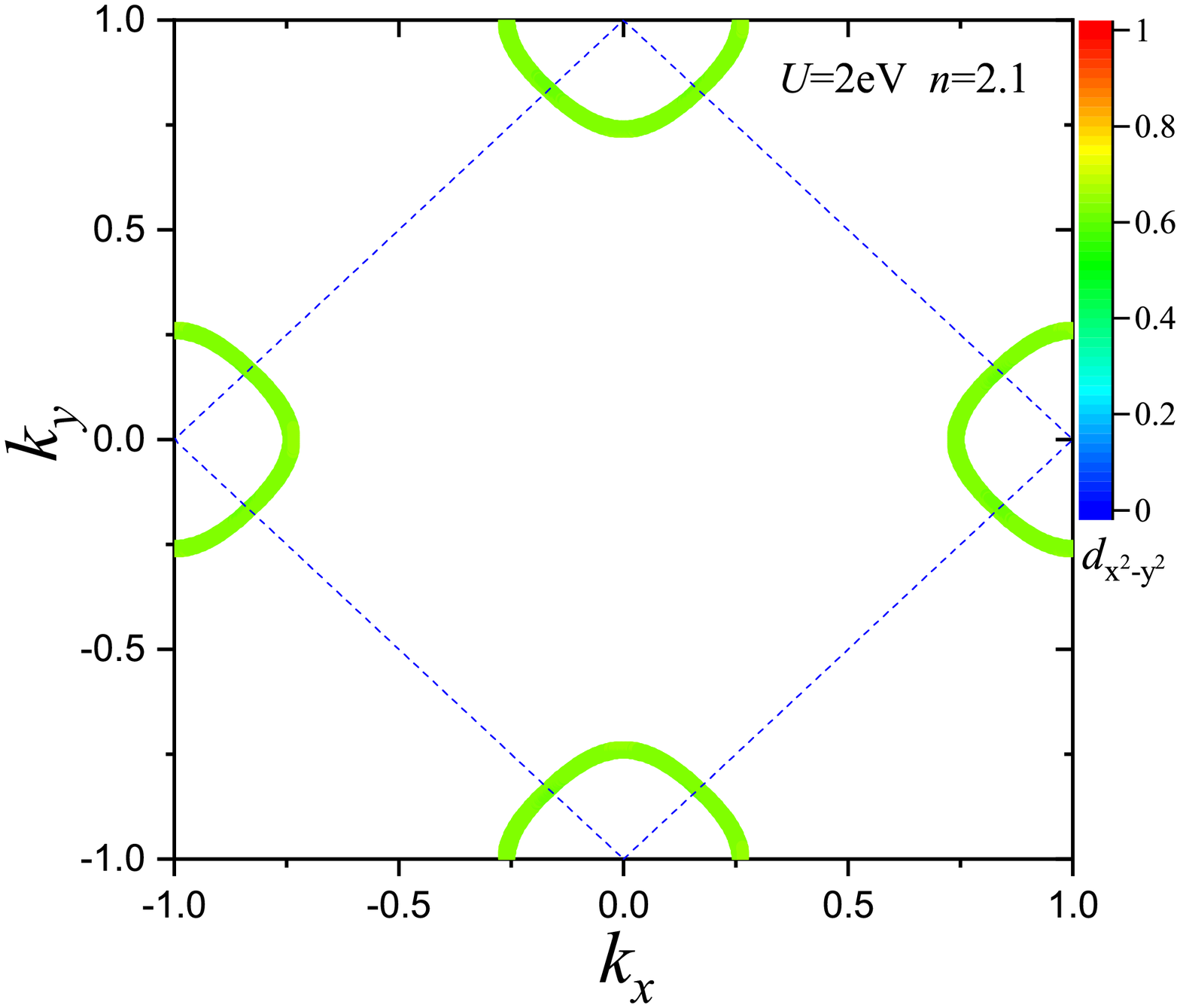}
\includegraphics[angle=0, width=0.49 \columnwidth]{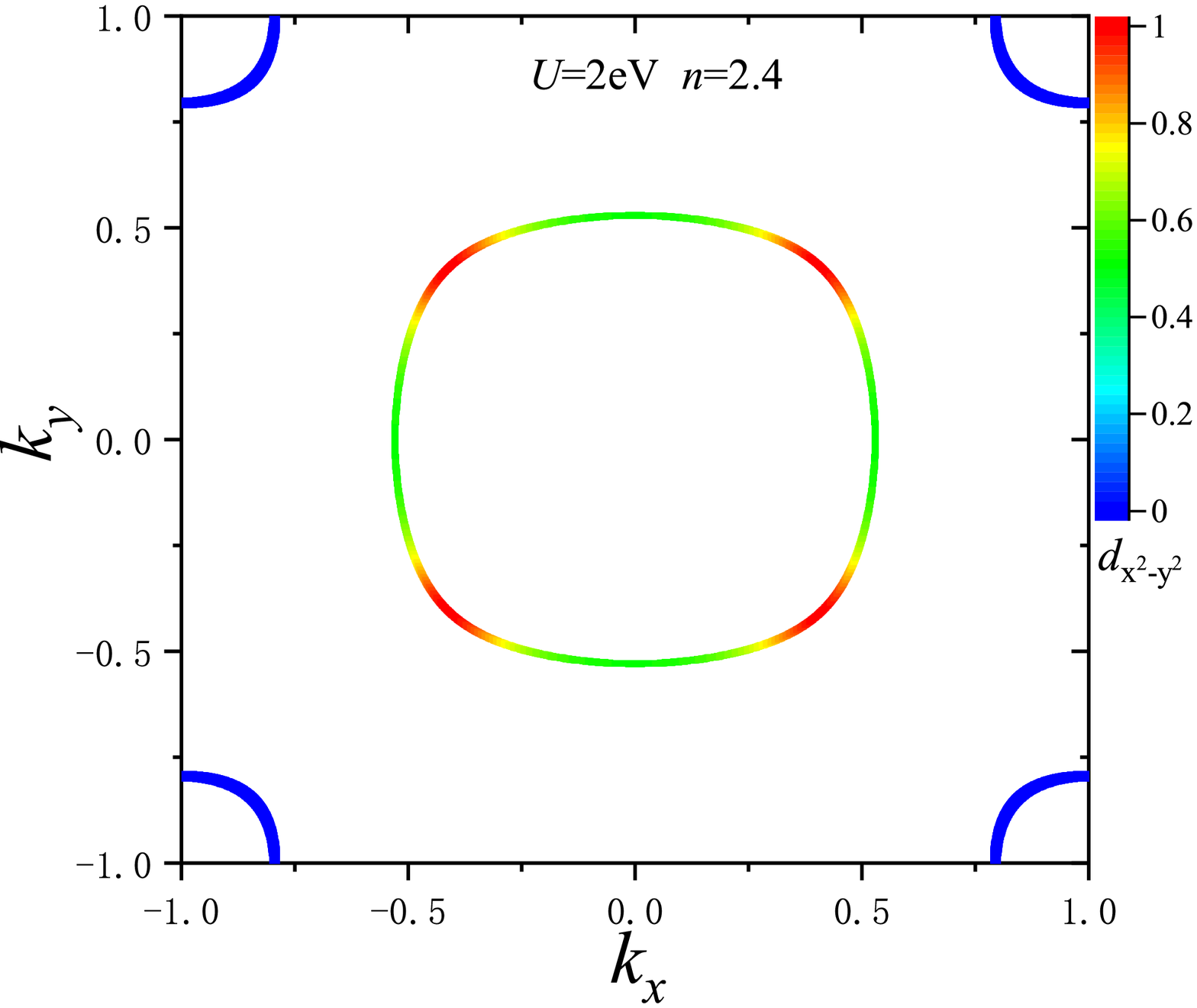} 
\includegraphics[angle=0, width=0.49 \columnwidth]{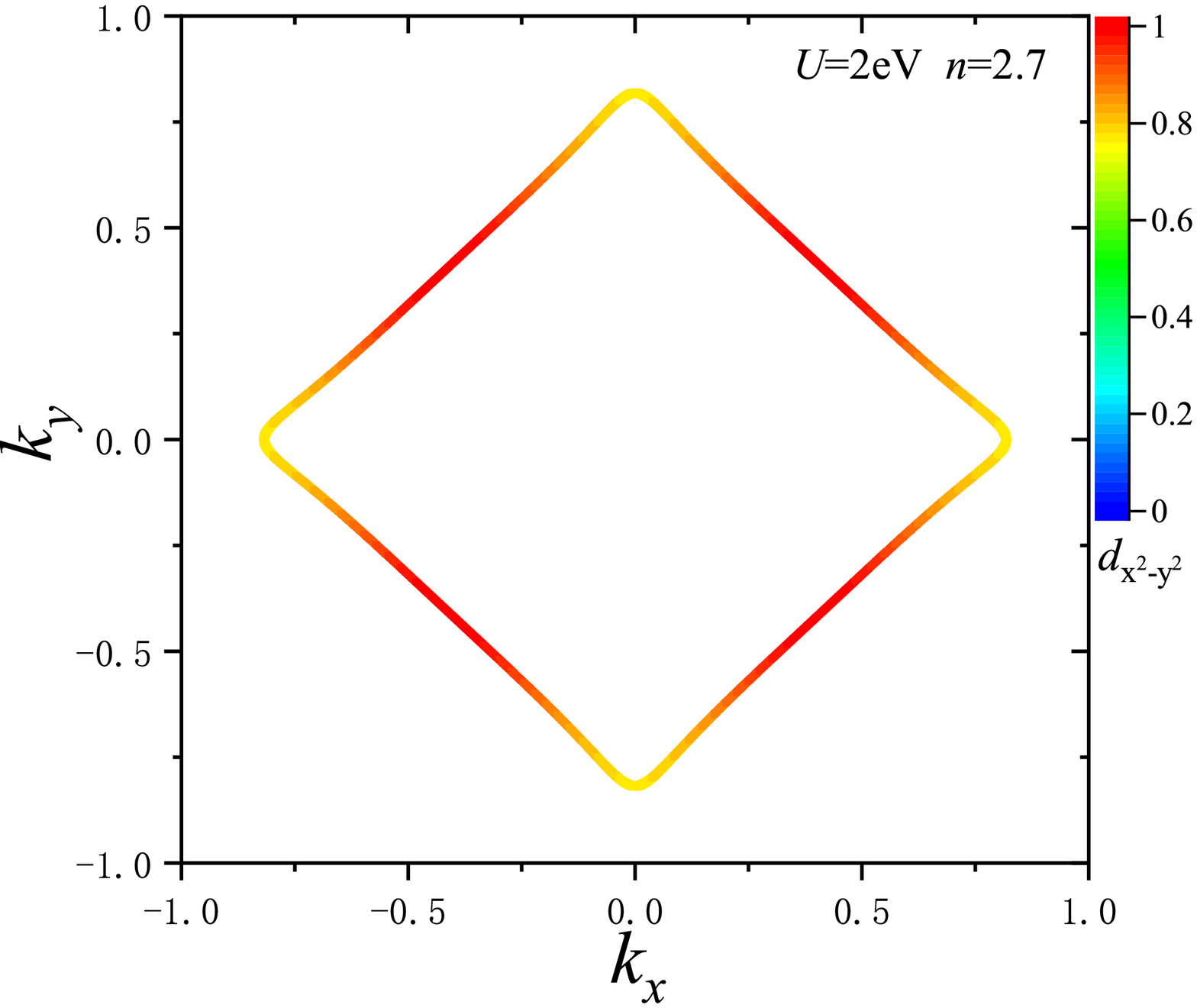}	 
\includegraphics[angle=0, width=0.49 \columnwidth]{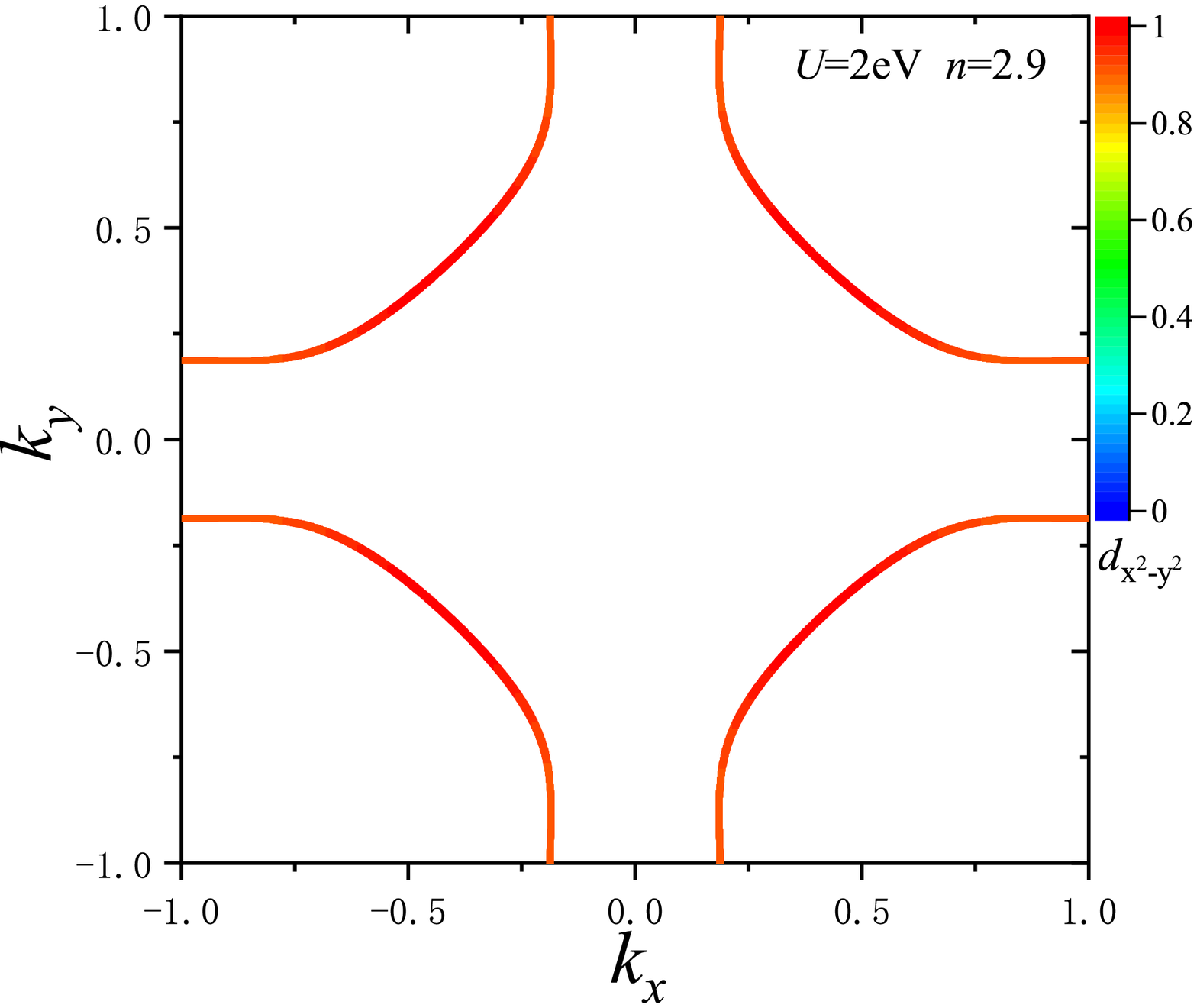} 	
 	\includegraphics[angle=0, width=0.49 \columnwidth]{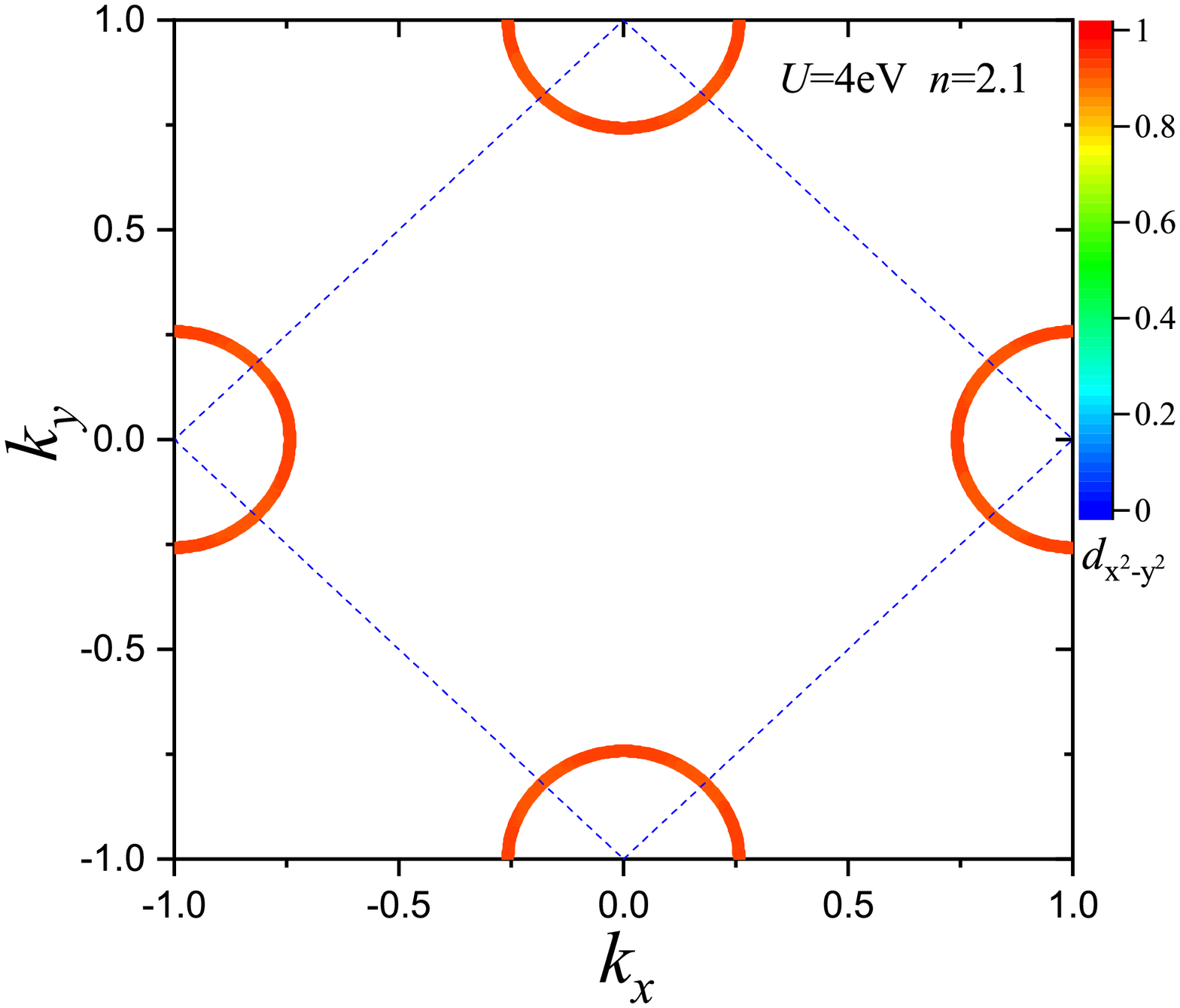}
 	\includegraphics[angle=0, width=0.49 \columnwidth]{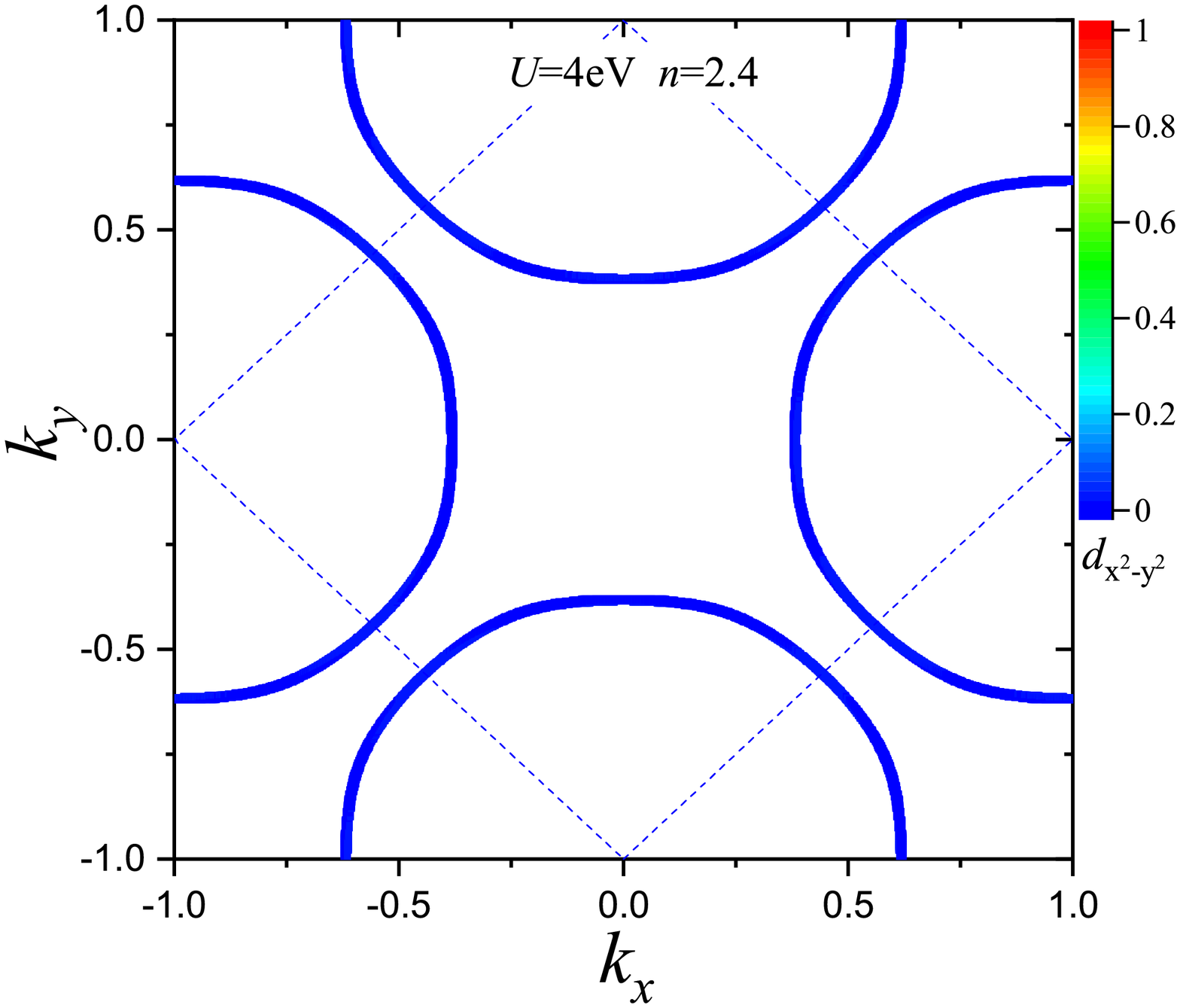} 
 	\includegraphics[angle=0, width=0.49 \columnwidth]{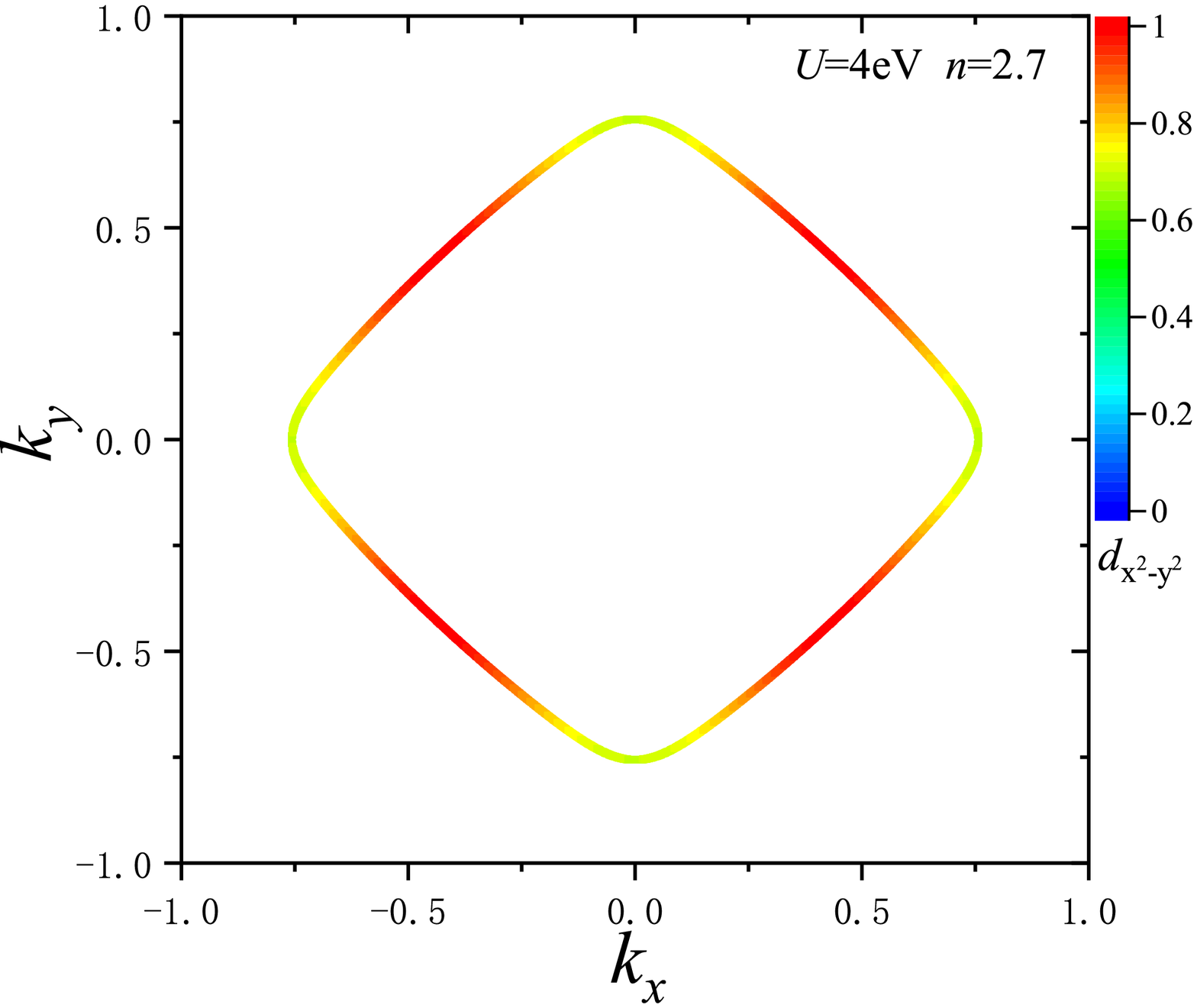}	 
 	\includegraphics[angle=0, width=0.49 \columnwidth]{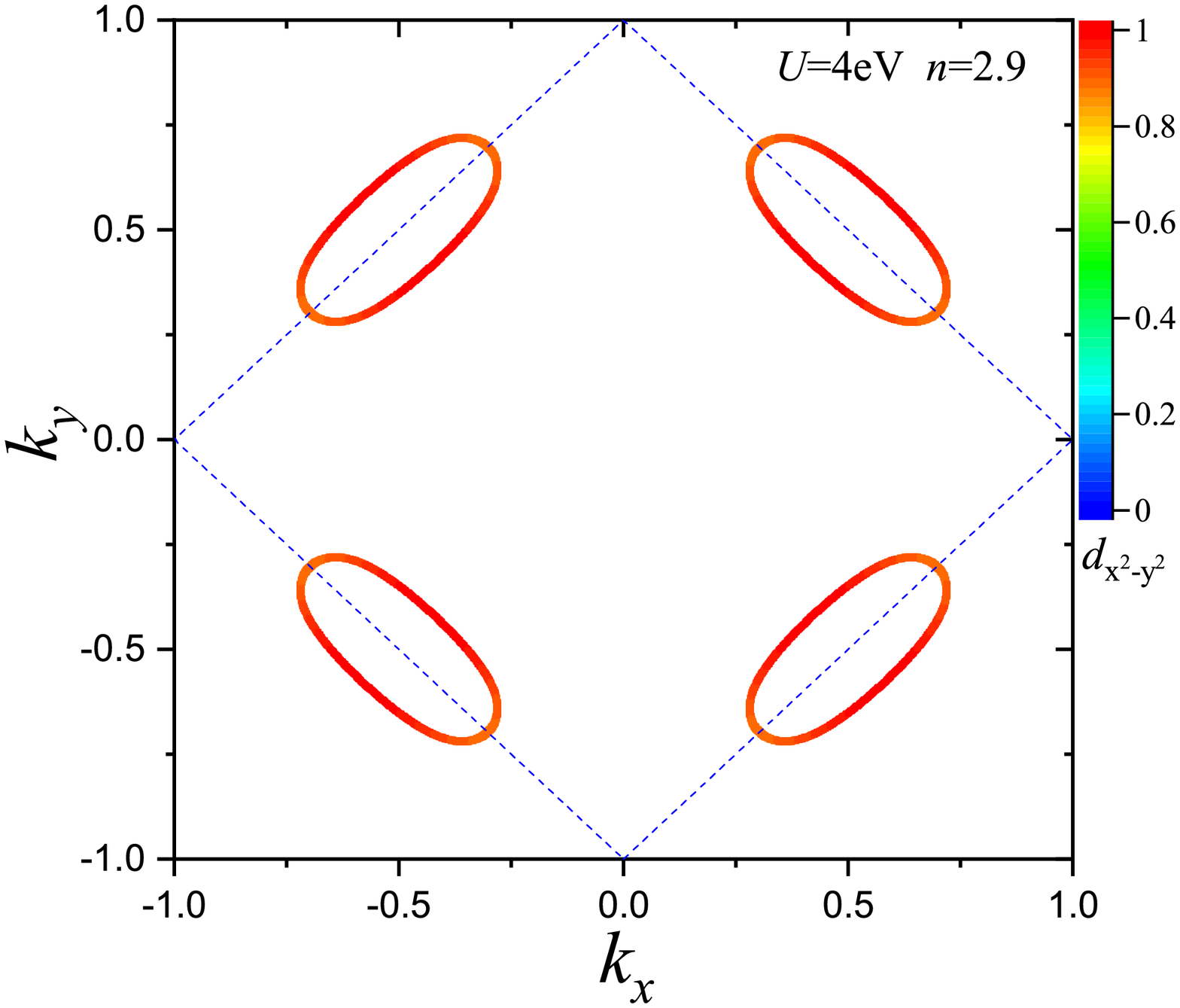}		
 	\caption{(Color online) The evolutions of the Fermi surfaces of Ba$_2$CuO$_{4-\delta}$ 
		with increasing doping at $U=2$ (upper two panels) and $U=4$ (lower two panels). 
		The color represents the orbital weight of the $d_{x^{2}-y^{2}}$ (0) and $d_{3z^{2}-r^{2}}$ (1)
		orbitals .}
 	\label{Fig.45}
 \end{figure}

We also obtain the evolution of the Fermi surfaces of Ba$_2$CuO$_{4-\delta}$ with increasing
filling number n at different correlation strengths. We plot the Fermi surfaces of the ground states 
of several typical filling numbers for U=2 eV in the upper panel and for U=4 eV in the lower panel
of Fig.\ref{Fig.45}. The Brilliouin zones with the dashlines indicate the folded zones due to the 
presence of the AF orders.
%
 %CCCCCCCCCCCCCCCCCCCCCCCCCCCCCCCCCCCCCCCCCCCCCCCCCCCCCCCC 
%%cccccccccccccccccccccccccccccccccccccccccccccccccccccccccccccccccccccccccccccc 
From Fig.\ref{Fig.45} we find that in the intermediate correlation regime of U=2 eV, with the
increase of filling number, the Fermi surfaces of the system  evolve from 4 small hole 
pockets in the corners at n=2.1 to both 4  small hole Fermi surfaces in the corners and a large 
electron pocket in the zone center at n=2.4, then to a more large electron pocket in the zone 
center at n=2.7 since the hole pockets are fully filled. The presence of two kind Fermi surfaces
is the direct consequence of two bands at n=2.1 and 2.4.

In the strong correlation regime with U=4 eV, the evolution of the Fermi surfaces with increasing
doping falls in expectation when $n<2.7$. It is interested that at n=2.9, due to the strong correlation,  
a large electron Fermi surface splits into four separated small pockets in the centers of the zone edge .

\section{Superconducting Pairing Strength}

To explore the unconventional superconductivity in Ba$_{2}$CuO$_{4-\delta}$, starting from  the original two-orbital
model Eqns.(1)-(5)  we also perform the numerical calculations for  the superconducting pairing strength  
arising from the spin-orbital fluctuations within the random phase approximation. 
We use the interaction parameters $U'=0.5U$ and $J_{H}=J_{P}=U/4$ when calculating the pairing strength.
The superconducting pairing strength of Ba$_{2}$CuO$_{3.2}$ as $U$ increasing and the pairing strength of 
Ba$_{2}$CuO$_{4-\delta}$ as hole doping $2.35<n<2.65$ at $U=2.0$ eV are presented in Fig.10. 

We find that the dominant pairing symmetry is mixed s+d-wave or d-wave in the parameter ranges 
we are interested. It shows 
that at n=2.6, the pairing strength $\lambda$ is negligible when the electronic 
correlation U is less than 1.6 eV; it becomes significantly with the s+d-wave pairing symmetry only when U $>$ 2 eV 
and critically increases with the d-wave pairing symmetry when U $>$ 2.2 eV, suggesting that 
superconducting  Ba$_{2}$CuO$_{4-\delta}$ is at least in the intermediate or strong correlated regime.
When U $>$ 2.4 eV, the magnetic susceptibility and superconducting pairing strength  diverge, more 
accurate methods, such as the fluctuation-exchange (FLEX) approximation, are expected.
%CCCCCCCCCCCCCCCCCCCCCCCCCCCCCCCCCCCCCCCCCCCCCCCCCCCCCCCC 
 \begin{figure}[htbp]
 	\centering
 	\includegraphics[angle=0, width=1.00 \columnwidth]{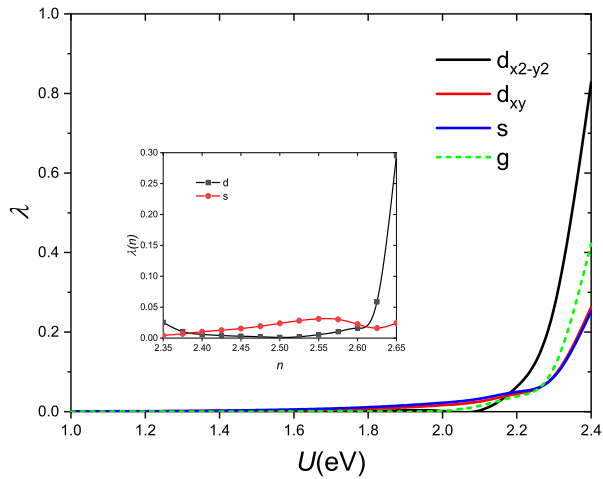}	
 	\caption{(Color online) The evolutions of the superconducting pairing strength $\lambda$ 
	        of Ba$_2$CuO$_{4-\delta}$ in the d-wave channel with the increasing correlations at n=2.6.
		Inset shows the dependence of the superconducting pairing strength $\lambda$ on increasing 
		doping at $U=2$. 
		}
 	\label{Fig.46}
 \end{figure}
%%cccccccccccccccccccccccccccccccccccccccccccccccccccccccccccccccccccccccccccccc 
From the doping dependence of the pairing strength shown in the inset of Fig.\ref{Fig.46} we find that  
at U=2 eV, the $\lambda$ does not monotonically vary with the electron filling number n. It shows
that when n$<$ 2.38 or n$>$2.61, the pairing symmetry is d-wave dominant, and the s-wave symmetry
becomes dominant  when 2.38$<$n$<$2.61. At the superconducting case n=2.6, 
%CCCCCCCCCCCCCCCCCCCCCCCCCCCCCCCCCCCCCCCCCCCCCCCCCCCCCCCC 
% \begin{figure}[htbp]
%  	\centering
%  	\includegraphics[angle=0, width=1.00 \columnwidth]{figure11-gap.eps}	
%  	\caption{(Color online) The superconducting pairing symmetry  
% 	        of Ba$_2$CuO$_{4-\delta}$ in the s-wave channel with the increasing correlations at n=2.6 
% 		and $U=2$. 
% 		}
%  	\label{Fig.47}
%  \end{figure}
%%cccccccccccccccccccccccccccccccccccccccccccccccccccccccccccccccccccccccccccccc 
the s-wave superconducting pairing strength becomes significant, meanwhile, the d-wave pairing weight
can not be neglected, suggesting a possible s+d-wave symmetry. These results demonstrate
that the superconducting pairing strength of Ba$_{2}$CuO$_{3.2}$ is significantly finite.
 
\section{Discussions and Conclusion}
\label{summary}

Therefore, regarding to the Scalapino {\it et al.}'s  two active orbiat model near the Fermi energy 
for recently discovered superconductor Ba$_{2}$CuO$_{4-\delta}$ with compressed octahedra,
we have studied the influences of the electronic correlation, magnetic correlation and doping evolution 
on the groundstate properties of Ba$_{2}$CuO$_{4-\delta}$ by utilizing the Kotliar-Runkenstein slave boson method. 
We demonstrate that at half filling, the correlated 
system displays two-band character with the  orbital nature of  $d_{3z^{2}-r^{2}}$ and $d_{x^{2}-y^{2}}$,
and transits from the PM to AFM, the Slater AFI, and Mott AFI 
when the electronic correlation U continuously increases from weak to strong. It is worthy noting that 
Ba$_{2}$CuO$_{3.5}$ is the Slater AFI phase when U$>$1.3 eV, or Mott AFI phase when U $>$ 4 eV.
At three-quarter filling, our study shows that the strongly correlated system exhibits single-band character
with the orbital nature of $d_{x^{2}-y^{2}}$, and transits from the PM to the Slater AFI, S+M AFI, 
and Mott AFI when the electronic correlation increases from weak to strong. 
This Mott AFI phase is in agreement with the AF Mott insulator obtained by using the $GGA+U$ method at large U 
for Ba$_{2}$CuO$_{3}$ \cite{XiangT-prm-3-044802-2019}.
With the increasing doping concentration from 2 to 3, due to the crystalline field splitting effect, the system evolves from 
two-band character with the  $d_{3z^{2}-r^{2}}$ and $d_{x^{2}-y^{2}}$ orbital nature to single-band one with the
$d_{x^{2}-y^{2}}$ orbital nature both for intermediate and for strong correlations.  
Due to the magnetic correlation, when n varies from 2 to 3, the magnetic phase diagrams in the intermediate correlation regime
and in the strong correlation regime are different: in the former, there exists only one AF phase when 2$<$n$<$2.17, 
in the latter the system displays first AF phase when 2$<$n$<$2.57 and second AF phase  when 2.76$<$n$<$3.
%
%The optimized doping superconducting phase Ba$_{2}$CuO$_{3.2}$ with n=2.6 lies 
%in the boundary between two-band and single-band case, implying that Ba$_{2}$CuO$_{3.2}$ is a candidate of 
%two-orbital or two-band superconductor, in agreement with the result by the Gutzwiller approach for a two-orbital
% Hubbard model \cite{Hu-j-p-prb-99-224515-2019}.
%
% Meanwhile, Ba$_{2}$CuO$_{3.5}$ and Ba$_{2}$CuO$_{3}$ are two distinct AF parent compounds,
%this indicates possible two kind superconducting pairing channels provided by exchanging spin fluctuations 
%in Ba$_{2}$CuO$_{4-\delta}$. 
Notice that we here assume the oxygen vacancy is homogeneously distributed, do not consider partial or full 
oxygen order, which has been considered in a few recent works \cite{XiangT-prm-3-044802-2019, Hu-j-p-prb-99-224515-2019}.

Regarding to concrete compound  Ba$_{2}$CuO$_{3.2}$,
from the preceding studies  one notices that determining the electronic correlation strength U is crucial
for understanding the properties of the normal and superconducting states of Ba$_{2}$CuO$_{3.2}$. 
This urges for more optical experiments, such as the X-ray absorption spectra (XAS) and resonant 
inelastic X-ray scattering, in high-quality Ba$_{2}$CuO$_{3.2}$ samples.
Up to date, despite two XAS experiments by Li {\it et al.} \cite{JinCQ-PNAS-116-12156-2019} and 
Fumagalli {\it et al.} \cite{Sala-physicaC-2021} provided some
information about the excited energy, a realistic U value remains unknown. Nevertheless, from the 
experimental spin coupling value about 150 meV  \cite{Sala-physicaC-2021}, comparing with the total-energy 
difference between the Neel AFM and paramagnetic phases shown in Fig.5, we estimate that U 
$\approx$  2.5 eV or so in Ba$_{2}$CuO$_{3.2}$, implying 
that Ba$_{2}$CuO$_{3.2}$ is at least an intermediated correlated system, or even a strongly correlated 
system. To exactly confirm the U value in Ba$_{2}$CuO$_{4-\delta}$ one needs to further perform the neutron 
scattering experiment to detect the magnetic moment per Cu spin and compare the theoretical magnetic 
moments in Fig.8. 

Meanwhile, it is interested whether Ba$_{2}$CuO$_{4-\delta}$ could be a two-dome  . 
when one takes U=2$\sim$ 2.5 eV for Ba$_{2}$CuO$_{4-\delta}$, our generalized magnetic  
phase diagrams in Fig.7, as well as the doping evolutions of the band structures in Fig.3 and of 
the Fermi surfaces in Fig.9, show that Ba$_{2}$CuO$_{3.2}$ is essentially a single-band superconductor, 
though this band consists of hybridized $d_{x^{2}-y^{2}}$ and $d_{3z^{2}-r^{2}}$ orbitals.  In this correlation regime,
the system behaves as single-band paramagnetic when n $>$ 2.2, or $\delta >$ 0.6, and has only one AFM 
parent phase, thus precludes the two-dome superconducting phases. 
Only in the strongly correlated regime with U $>$ 4 eV, can the system display two different AFM parent 
phases at n=2 and 3, corresponding to two-band and single-band nature, respectively. In this situation, different 
type spin fluctuations contribute different superconducting pairing strengths, suggesting the possibility of 
two-dome superconducting phases. 
A few of recent theoretical works suggested Ba$_{2}$CuO$_{4-\delta}$ is weak correlated 
\cite{Maier-prb-99-224515-2019}, or strong correlated \cite{XiangT-prm-3-044802-2019,Hu-j-p-prb-99-224515-2019}; 
we expect further experiment could finally resolve this issue.

%ccccccccccccccccccccccccccccccccccccccccccccccccccccccccccccccccccccccccccccccccccccccccccccccccccccc

\begin{acknowledgements}

This work was supported by the National Natural Science Foundation of China under Grant nos.
$11774350$  and $11534010$, 
The calculations were performed in Center for Computational Science of CASHIPS, 
and partly using Tianhe-2JK computing time award at the 
Beijing Computational Science Research Center(CSRC).

\end{acknowledgements}

%\nocite{*}
%\bibliography{apssamp}

\end{document}